\documentclass[twocolumn]{aastex631}

\newcommand{\xie}{\color{black}}
\newcommand{\xx}{\color{black}}
\newcommand{\xxx}{\color{black}}
\newcommand{\Su}{\color{black}}
\newcommand{\jw}{\color{black}}

\usepackage{color}

\received{April 23, 2021}
\revised{August 27, 2021}
\accepted{September 28, 2021}
\submitjournal{AJ}

\shorttitle{Demographics of Exoplanets in Binaries (DEB-I)}
\shortauthors{Su et al.}

\graphicspath{{./}{figures/}}

\begin{document}

\title{Demographics of Exoplanets in  Binaries (DEB). I. Architecture of S-Type Planetary Systems Revealed by the RV Sample.}

\correspondingauthor{Ji-Wei Xie}
\email{jwxie@nju.edu.cn}
\author{Xiang-Ning Su}
\author{Ji-Wei Xie}
\author{Ji-Lin Zhou}

\affiliation{School of Astronomy and Space Science, Nanjing University, Nanjing 210023, People’s Republic of China}
\affiliation{Key Laboratory of Modern Astronomy and Astrophysics, Ministry of Education, Nanjing 210023, People’s Republic of China}
\author{Philippe Thebault}
\affiliation{LESIA, Observatoire de Paris, Meudon Principal Cedex, France}

\begin{abstract}

Although the sample of exoplanets in binaries has been greatly expanded, the sample heterogeneity and observational bias are obstacles toward a clear figure of exoplanet demographics in the binary environment. 
To overcome the obstacles, we conduct a statistical study that focuses on S-type (circumstellar) planetary systems detected by the Radial Velocity (RV) method. 
We try to account for observational biases by estimating, from available RV data, planet detection efficiencies for each individual system. 
Our main results are as follows: 
(1) Single (resp. multiple) planetary systems are mostly found in close (wide) binaries with separation $a_B< (>)\sim$ 100-300 AU. 
(2) In binaries, single and multiple-planet systems are similar in 1-D distributions of mass and period as well as eccentricity (in contrast to the ``eccentricity dichotomy" found in single star systems) but different in the 2-D period-mass diagram. 
Specifically, there is a rectangular-shaped gap in the period-mass diagram of single-planet systems but not for multiples. 
This gap also depends on binary separation and is more prominent in close binaries. 
(3) There is a rising upper envelope in the period-mass diagram for planets in wide binaries as well as in single stars but not in close binaries. 
More specifically, there is a population of massive short period planets in close binaries but almost absent in wide binaries or single stars. 
We suggest that enhanced planetary migration, collision and/or ejection in close binaries could be the potential underlying explanation for these three features.

\end{abstract}


\section{Introduction} \label{sec:intro}
{\xie
Stars are thought to be commonly born and found in binary/multiple systems \citep{1991A&A...248..485D, 2010ApJS..190....1R}.
Therefore, the demographics of exoplanets in binaries play a crucial role in statistically studying the whole exoplanet population in our Galaxy.
Furthermore, the diverse orbital configurations and rich dynamics in planet-bearing binary systems provide valuable conditions to test various theories and models of planet formation and evolution \citep{2015pes..book..309T,2019Galax...7...84M}.

There are generally two orbital configurations of planets in binaries \citep{1982OAWMN.191..423D}.
One is called P-type, where planets orbit both the  binary stars, i.e., circumbinary planets, e.g., Kepler 413 ABb \citep{2013ApJ...770...52K}.
The other is called S-type, where planets orbit one of the binary stars, i.e., circumstellar planets, e.g., $\gamma$ Cephei Ab \citep{1988ApJ...331..902C,2003ApJ...599.1383H}. 
Most planets in binaries are currently found in S-type.
At the time of writing this paper ({\Su04-March-2021}), by combining the two catalogues (see section 2.1 for details) retrieved from the Extrasolar Planets Encyclopaedia and the Catalogue of Exoplanets in Binary Star Systems \citep{2016MNRAS.460.3598S}, there are {\Su211} confirmed S-type planets. 
Among them, {\Su 116} are from radial velocity (RV) surveys (e.g., $\gamma$ Cephei Ab \citep{1988ApJ...331..902C,2003ApJ...599.1383H}, Tau Bootis b \citep{1997ApJ...474L.115B}), {\Su 89} from transit surveys (e.g., {\Su Kepler-68 b \citep{2013ApJ...766...40G}}, HD 202772 b  \citep{2019AJ....157...51W}) and 6 from others, e.g., microlensing surveys \citep{2016AJ....152..125B,2014Sci...345...46G}, and {\Su direct imaging surveys (e.g. 51 Eri \citep{2015ApJ...813L..11M})}.
There are only 28 confirmed P-type planets, and a large fraction of them are from transit surveys, e.g., Kepler-16AB b \citep{2011Sci...333.1602D} and Kepler-47 b,c and d \citep{2012Sci...337.1511O,2019AJ....157..174O} etc.
Note, the above numbers only reflect a small but relatively well characterized portion of  planet-bearing binary systems.
Many candidate planet-bearing binary systems, e.g., those from imaging followup observations of Kepler planet host stars \citep{2014ApJ...791..111W,2014ApJ...783....4W, 2016AJ....152....8K, 2017AJ....153...66Z,2017AJ....153...71F,2018AJ....155..161Z} are currently not included in the above two catalogues because either the planetary nature or stellar binarity remains to be confirmed.

When mining the observational data, there are two commonly used approaches to study the effect of stellar binarity on planetary systems.
On the one hand, by calculating the binary fraction of planet hosts and comparing it to those of non-planet hosts or field stars, one can learn whether/how planet formation is suppressed by binary stars \citep{2007A&A...468..721B,2020Galax...8...16B,2011IAUS..276..409E,2014ApJ...791..111W,2014ApJ...783....4W,2016AJ....152....8K,2017AJ....153...66Z,2018AJ....155..161Z,2018AJ....156...31M,2019arXiv191201699M,2019MNRAS.485.4967F,2021arXiv210112667F}.
{\jw One difficulty with this approach is to remove or at least quantify the effects of observational bias against finding planets in binaries. 
A major bias often comes from the target selection process itself, which is severely biased against binaries in most RV surveys, but also significant for transit surveys, e.g, Kepler \citep{2021AJ....161..231W}.}
On the other hand, by investigating the distributions  of planet properties (e.g., mass, period, and etc.) in known planet-bearing binary systems, one can learn how planetary architecture is sculpted by binary stars, e.g., \citep{2012A&A...542A..92R}.
{\jw It is the latter approach that we adopt in this paper. Furthermore, we only focus on S-type planet systems from RV surveys to reduce the sample complexity and thus better take into potential observational bias.
Since we just care about `relative' properties (e.g., period-mass distribution) rather than  `absolute' properties (e.g, absolute occurrence rate), our analysis is not affected by binary-hostile target selection biases in RV surveys. The main potential bias we need to correct for is the varying planet detectability performances depending on planetary mass and period, which could over-represent (or under-represent) some types of planets in our sample. We could also have, for the same location in the planetary mass-period diagram, different detectabilty thresholds depending on the considered system.
For example, some systems may have poorer RV measurements due to large stellar activity or light contamination of close binary stars,  which would thus bias against finding smaller planets on longer periods in these systems.
}

{\xxx 
Some previous statistical studies have already explored the properties of S-type planets in RV surveys, but for more limited samples}. 
With a small sample of 9 S-type planets, \citet{2002ApJ...568L.113Z} found that planets in binaries seem to follow a different period-mass distribution than planets in single star systems.
Increasing the sample size to 19, \citet{2004A&A...417..353E} confirmed this trend by pointing out that the few most massive ($Msini>2 M_{\rm Jup}$) short-period ($P<40$ days) planets are exclusively in binary systems.
Doubling the sample size to $\sim$40, \citet{2007A&A...462..345D} found that the above mass difference is significant only in tight binaries (binary separation, $a_B < 100-300$ AU).
\citet{2010ApJ...709L.114D} then found that planets in tight binaries ($a_B \lesssim 100$ AU) are significantly more massive than those in wide binaries ($a_B \gtrsim 100$ AU).
\citet{2012A&A...542A..92R} compiled a larger sample of planets in binaries and found that (1) no planets are in binaries with a projected separation of less than 10 AU (2) multi-planets systems are exclusively in wide binaries ($a_B>100$ AU) and (3) planetary mass decreases with an increasing binary separation.

In this paper, which is the first (Paper I) of a series of studies on the Demographics of Exoplanets in Binaries (DEB for short), we focus on the architecture of S-type planetary systems detected by the RV method.
With the significantly expanded sample that we compile ({\Su110} S-type RV planets), we aim to (1) finding new patterns and (2) revisiting the above reported patterns, e.g., the period-mass distribution, by taking into account the RV detection efficiency in detail. 

This paper is organized as the follows.
In section \ref{sec:method}, we describe our planet sample and the method used to consider the effect of the RV detection efficiency.
In section \ref{sec:results}, we present our results.
Finally, we discuss and summarize the paper in sections \ref{sec:discussion} and \ref{sec:summary}.
}

\begin{figure}
\plotone{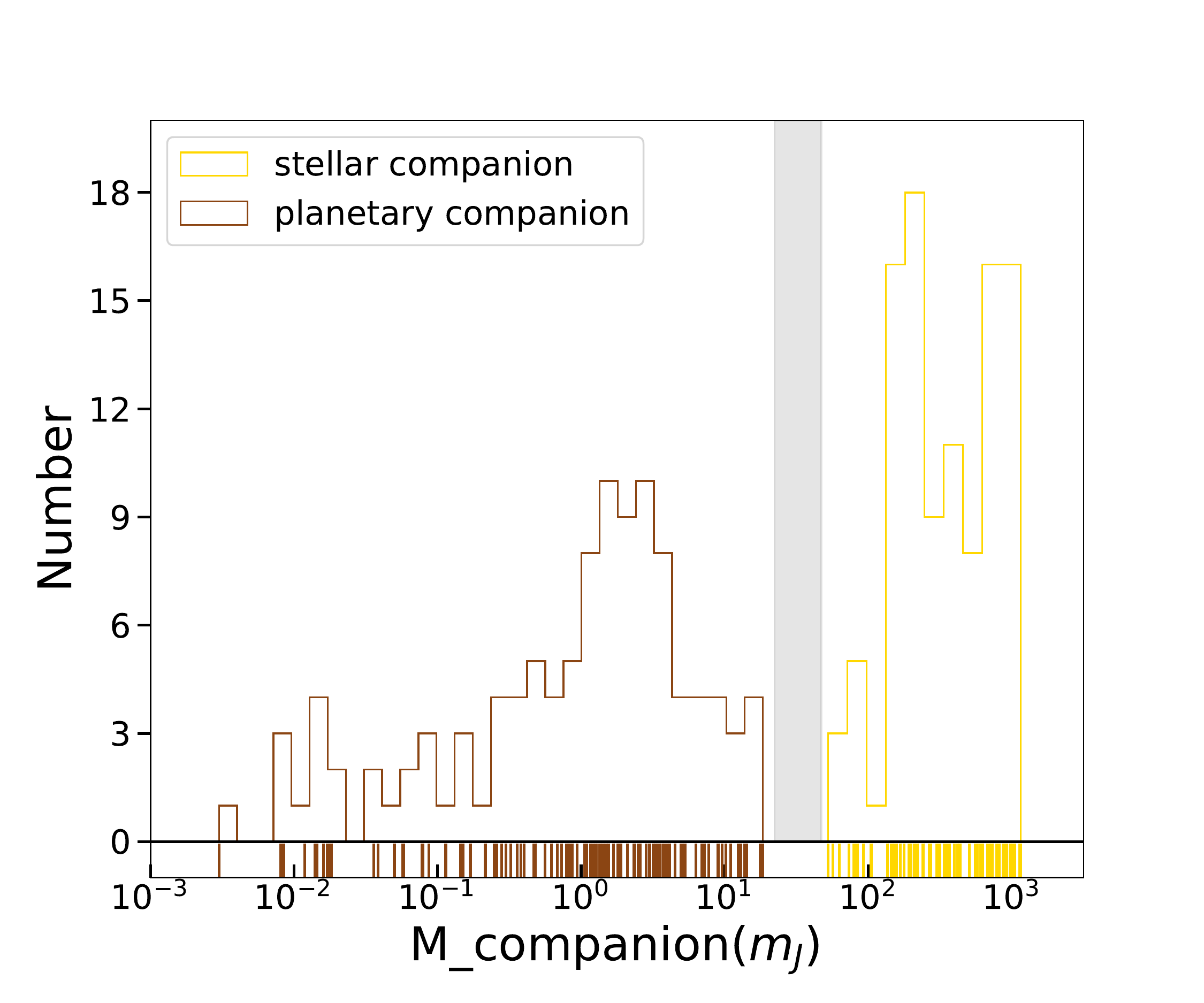}
\caption{{\Su 
The mass (or Msini) histogram of all companion objects in our sample.
Apparently, there is gap (grey shaded) between 20 and 50 Jupiter masses, which separates planetary objects from stellar objects.
}
\label{fig:m_comp}}
\end{figure}

\begin{figure}
\plotone{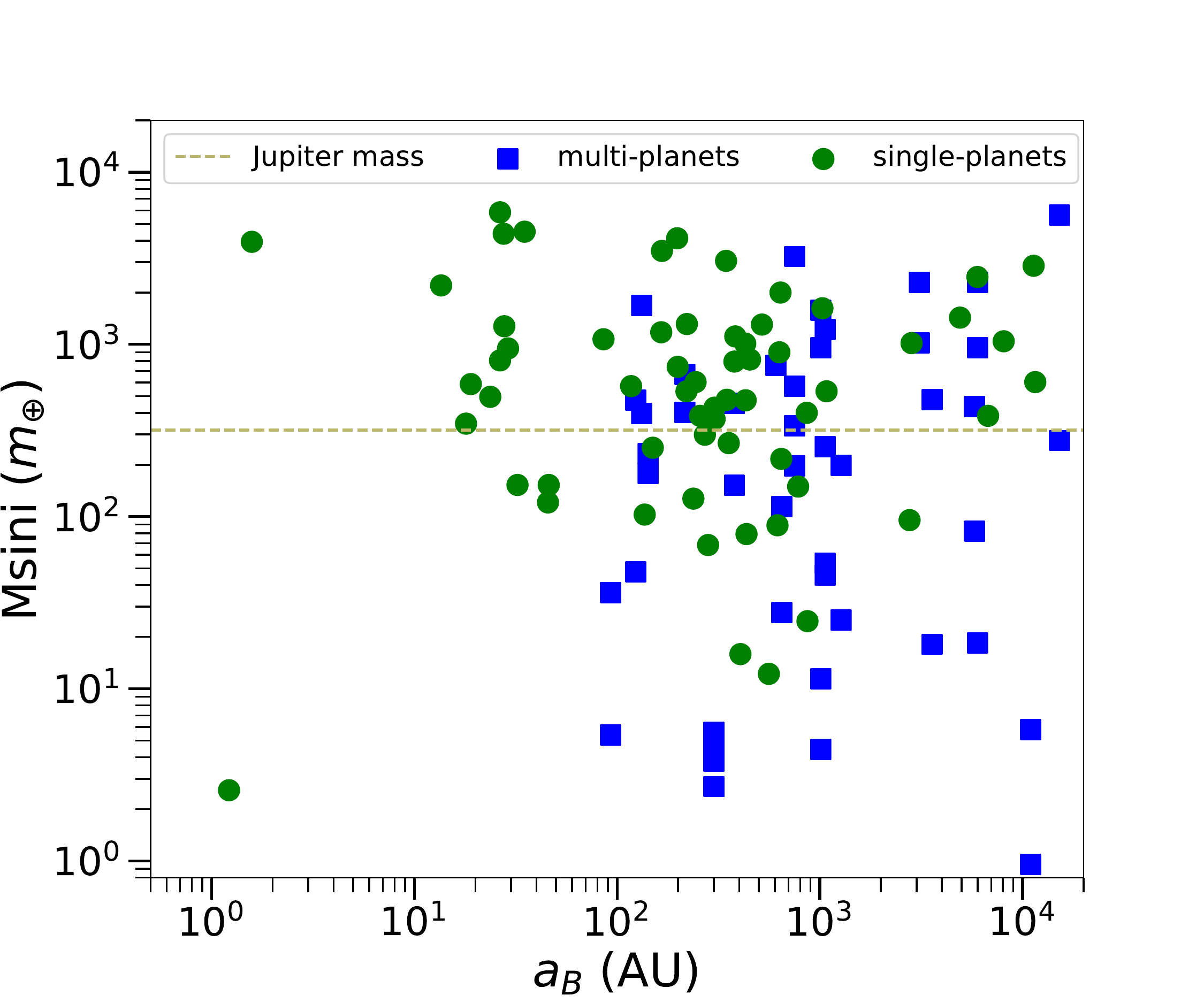}
\caption{{\xie 
Overview of our sample:  Planet mass as a function of  binary separation. 
All the single-planet systems are represented by green circles, while all planets in multiple-systems are represented by blue squares.
The light yellow dash line indicates the Jupiter mass.
}
\label{fig:aB-mp}}
\end{figure}

\begin{figure*}
\begin{center}
\includegraphics[width=0.95\textwidth]{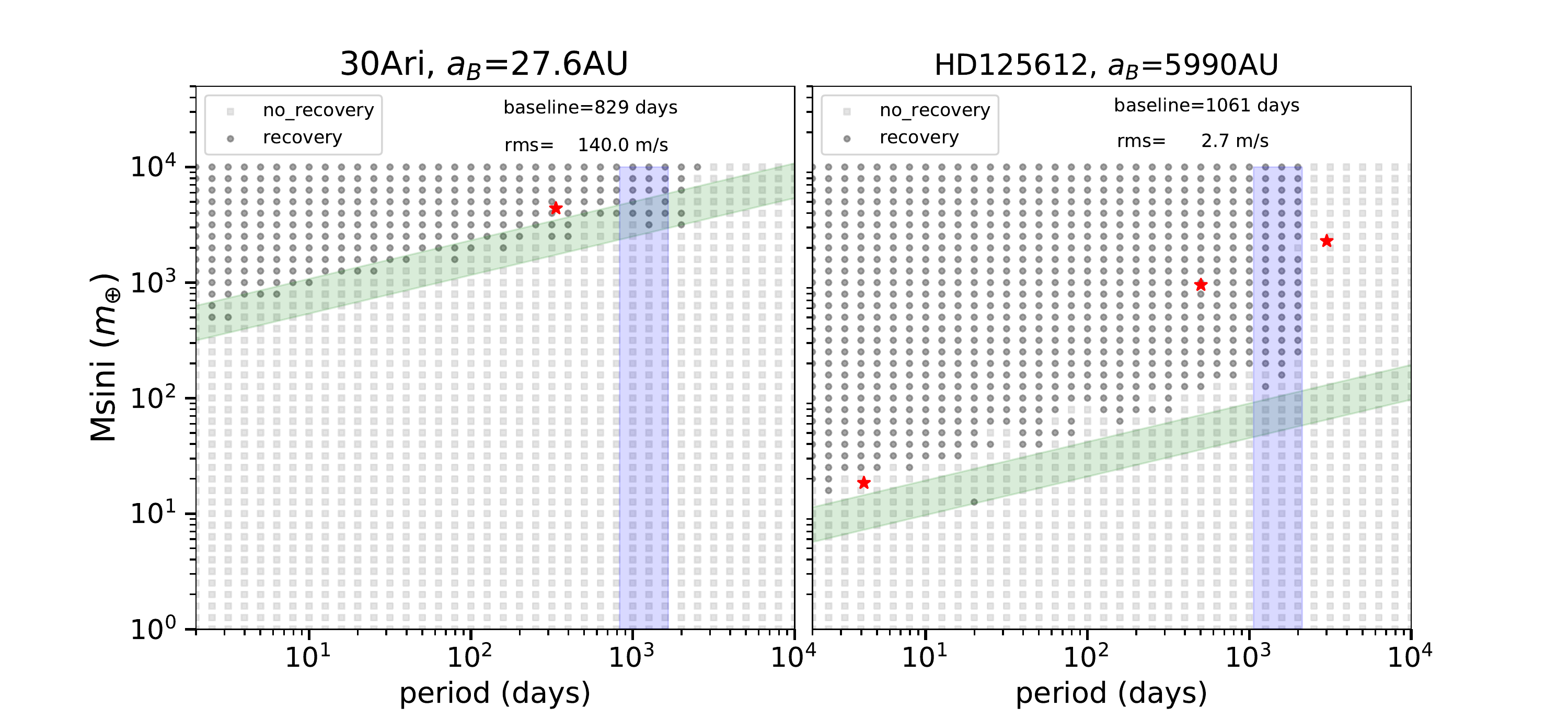}
\end{center}
\caption{{\xie 
Two examples of RV signal injection-recovery test. 
30 Ari is a close binary system with one planet (red star in the left panel), and HD 125612 is wide binary with three planets (red stars in the right panel). 
The filled  dark  grey circles (light squares) show  where  the hypothetical  planets  can (can not) be  recovered  from  the  simulated  RVs. 
The green shaded regions mark where $RV_{signal}=(1-2)\times rms$.
The purple shaded regions mark where  $period= (1-2)\times the \ time \ baseline$ of the simulated RV data.
}
\label{fig:example}}
\end{figure*}

\section{Sample and Method} \label{sec:method}
\subsection{Sample} \label{subsec:samples}

{\jw
We build our RV planet sample by  retrieving data (on 04-March-2021) from  two online catalogues. i.e., (1)  the catalog of exoplanet-hosting binaries with separations up to 500 AU \footnote{\url{http://exoplanet.eu/planets_binary/}}, (2) the catalogue of exoplanets in binary star systems \citep{2016MNRAS.460.3598S}\footnote{\url{https://www.univie.ac.at/adg/schwarz/multiple.html}}.
Specifically, we first retrieved all RV planet systems in the first catalog, which is almost exhaustive for all binaries of separation less than $\sim 500$ AU. Then, we turned to the second catalog, which provided most planets in binaries with separation from $\sim$500 up to $\sim$15000 AU in our sample. 
We note that HD 7449 c is reported as planet with minimum mass of 19 $m_{J}$ in the second catalog but not recognized in the first catalog. 
In fact, HD 7449 c is only inferred from the long term RV trend which could be due to the stellar companion \citep{2016ApJ...818..106R}.
Therefore, we removed this unconfirmed planet and thus HD 7449 is included as a single planet system in our sample.}
{\jw Furthermore, in order to have a unified criteria to define our sample, we adopted an upper mass limit (20 $m_{J}$) for planetary objects and a lower mass limit ($50 m_{J}$) for stellar objects. 
This criterion is motivated by the mass (or msini) distribution of the companion objects in Figure 1, which shows a gap between 20 and 50 Jupiter masses.
Under this definition, HD 87646 c and HD 4113 c were treated as stellar objects instead in our sample. 
}
In addition, systems HD 41004 AB, HD 20782/HD 20781 and HD 133131 AB have RV planets orbiting both the primary and companion stars, and they are treated as two single S-type planet systems in our analyses. 
Systems HD 80606, HD 20781, HD 103774 and HD 93385, whose RV data are not publicly available, are thus excluded from our analyses. 

With these criteria, we obtain {\Su 110} planets in {\Su 80} systems, {\Su61 (19)} of which are single (multiple) planet systems. 
Table \ref{tab:tab1} lists the properties of these planets in our sample.
Figure \ref{fig:aB-mp} is an overview of our sample in the plane of Msini-$a_B$, where Msini is the minimum mass of planet and $a_B$ is {\Su the semi-major axis of binary orbit}. 
{\Su  Our samples also include 23 planets in 16 triple star systems and quadruple star systems. 
In these cases, $a_B$ reported in Table \ref{tab:tab1}  should be treated as the semi-major axis of the closest planet-host companion.}


\begin{figure*}
\begin{center}
\includegraphics[width=0.99\textwidth]{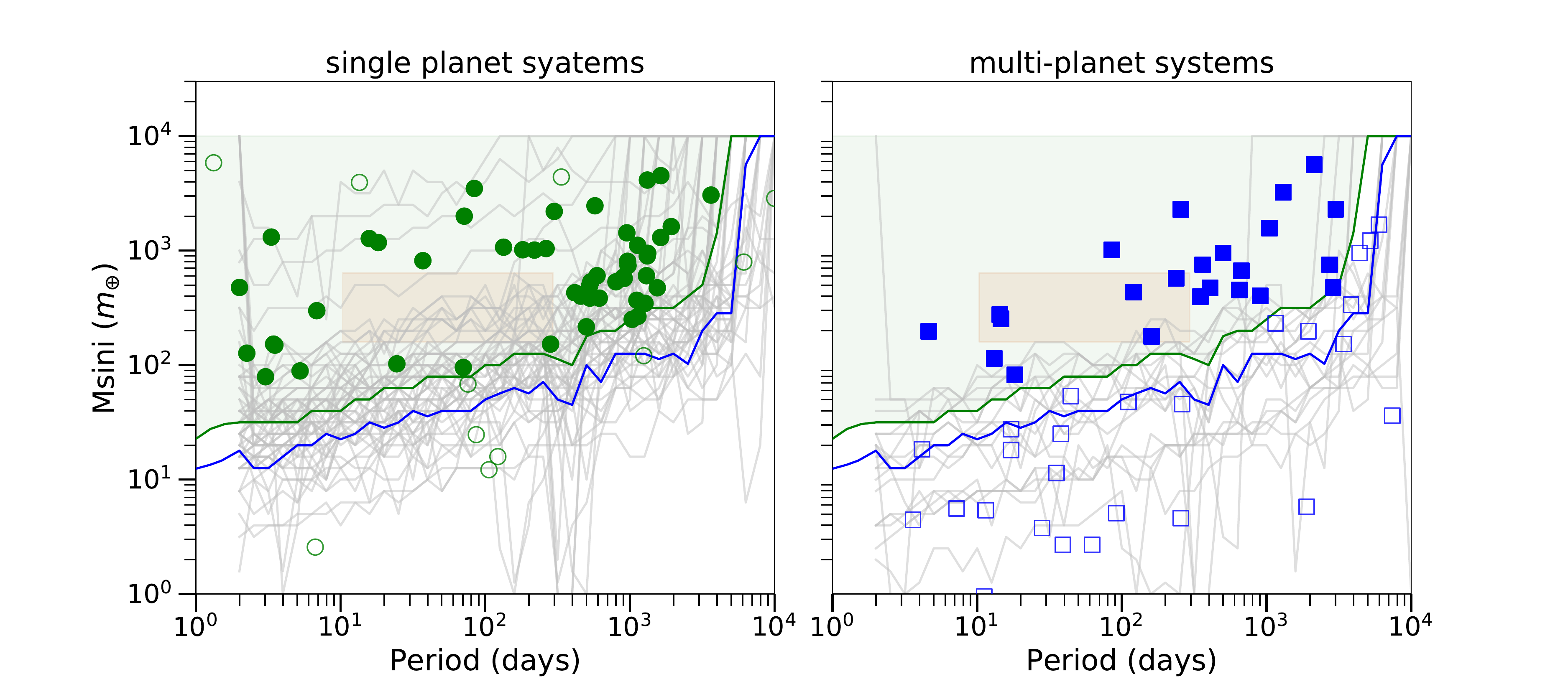}
\end{center}
\caption{ {\xie 
Period-mass diagrams for single-planet systems (green circles in the left panel) and multiple-planet systems (blue squares in the right panel). 
The green and blue curves represent the medians of the detection limits (grey curves) of singles and multiples, respectively. 
Systems with RV rms larger than $100$ ms$^{-1}$ or planets below the single median detection limit (green curves) are marked with open symbols.
The shaded rectangle marked the gap, a planet desert in the singles (left panel, see more in the text in section 3.1.3). 
}  
\label{fig:s_m_pmp}}
\end{figure*}

\subsection{Quantifying the RV detection constraints} \label{subsec:method}

{\xx Different systems could differ significantly in the precision of their RV measurements, causing an observational bias that systems with poorer RVs bias against finding smaller planets with longer period. 
In order to quantify this bias, we perform the following RV signal injection-recovery test in each system of our sample.}

{\xie
First, we collect the RV data of each system and extract the RV residual, $RV_{residual}$, by subtracting the observed planet signals using the  RV fitting tool {\it Systemic2} (\url{http://www.stefanom.org/console-2/},  \citep{2009PASP..121.1016M}).
{\jw During the above RV fitting process, we also removed the long term RV trend from binary orbits in some systems (e.g., Gliese 15, gamma Cephei, tau Boot A, HD41004 A/B, HD196885).
For a few systems (HD7449, HD 8673, HD59686, HD87646, HD2638, HD 30856, HD 42936, HD126614 and 30 Ari), the $RV_{residual}$ were extracted directly from the figures (converting graph into numbers) in the corresponding planet discovery papers because the RV data were not readily available.
}

Next, we inject a test RV signal, i.e., $RV_{signal}$ of a hypothetical planet with a given minimum mass ($Msini$) and orbital period (P).
Ignoring the effect of orbital eccentricity, we have
\begin{equation}
   RV_{signal} = \sqrt[3]{\frac{2\pi G}{P}\frac{(Msini)^3}{(M_*+M)^2}}\cos(\omega+\frac{2\pi t}{P}),
   \end{equation}
where $M_*$ is the mass of host star, $M$ is the mass of hypothetical planet,
$G$ is  gravitational constant,
$i$ is the inclination between the normal of the planet's orbit and the line of sight,
$\omega$ is the argument of periastron (drawn randomly from 0 to $2\pi$) of the planet's orbit, and $t$ is observing time. 

Then, we generate a set of simulated RV data, 
\begin{equation}
RV_{sim}=RV_{signal}+RV_{residual}+RV_{error},
\label{RV}
\end{equation}
 by combining the test planet signal with the RV residual and the RV measurement error.
 Here, $RV_{error}$ is a random number that follows the normal distribution, i.e., $N(0,\sigma_{RV})$, where $\sigma_{RV}$ is the reported RV uncertainty. 

Finally, we evaluate whether the injected planet signal can be recovered from the simulated RV data.  
The criteria are (1) the peak location of the periodogram \citep{1976Ap&SS..39..447L,1982ApJ...263..835S} should be within $25\%$ away from the orbital period of the injected planet, i.e., $|P_{peak}-P_{inject}|<0.25*P_{inject}$ and (2) the False Alarm Probability (FAP) of the  periodogram peak should be less than $1\%$.
The FAP is calculated via a bootstrap method \citep{2010ApJS..191..247T}. 
Specifically, we randomly shuffled the simulated RV data from Equation \ref{RV}, and record the power of the periodogram peak \citep{1976Ap&SS..39..447L,1982ApJ...263..835S}. 
We repeat such a random calculation 1000 times, and the FAP is set as the fraction of times when the record power is larger than the original one. 

Figure \ref{fig:example} shows the results of the RV signal injection-recovery tests for two individual systems (30 Ari and HD 125612, see also in Figures \ref{fig:figa0}-\ref{fig:figa3} in the Appendix for the results of all systems of our sample).
For each system, the RV signals are generated by hypothetical planets which are uniformly distributed in the log(P)-log(Msini) plane.
As can be seen, the filled dark grey  circles show where the hypothetical planets can be recovered from the simulated RV data. 
The bottom envelopes of the recovered region generally follow the line where $RV_{signal}=1-2 \times rms$, and there is a long period cut-off due to the time baseline of the RV data.
On the left is 30 Ari, a close binary ({\Su$aB=27.6$ AU}) hosting a single massive planet candidates ($Msini = 13.82 M_J$), while on the right is HD 125612, a wide binary ({\Su$aB=5990$ AU}) hosting three planets with masses of $Msini = 7.2, \,3.0,\, 0.058 M_J$.
{\xxx Figure \ref{fig:example} illustrates a potential detection bias of the RV method. Indeed, among the three planets of HD 125612, only the most massive one is close to the recovery region of 30 Ari, while the other two less massive planets would not have been discovered even if they were truly existing in the 30 Ari system.
{\jw This could be due to the fact that, because RV precision is generally poorer in close binaries (due to light contamination or inappropriate stellar properties e.g., large $vsini$)}, only the signal of the largest planets of these systems can be retrieved, leaving potential additional smaller planets undetected (whereas such planets would have been detected in wider binaries or single stars). As a consequence, the fraction of single-planet systems could be artificially overestimated in close binaries.
To address this potentially crucial issue, we will have to take into account the statistical average of detection limits for our respective samples of single-planet and multiple-planet systems}.
}

\begin{figure*}
\begin{center}
\includegraphics[width=1.0\textwidth]{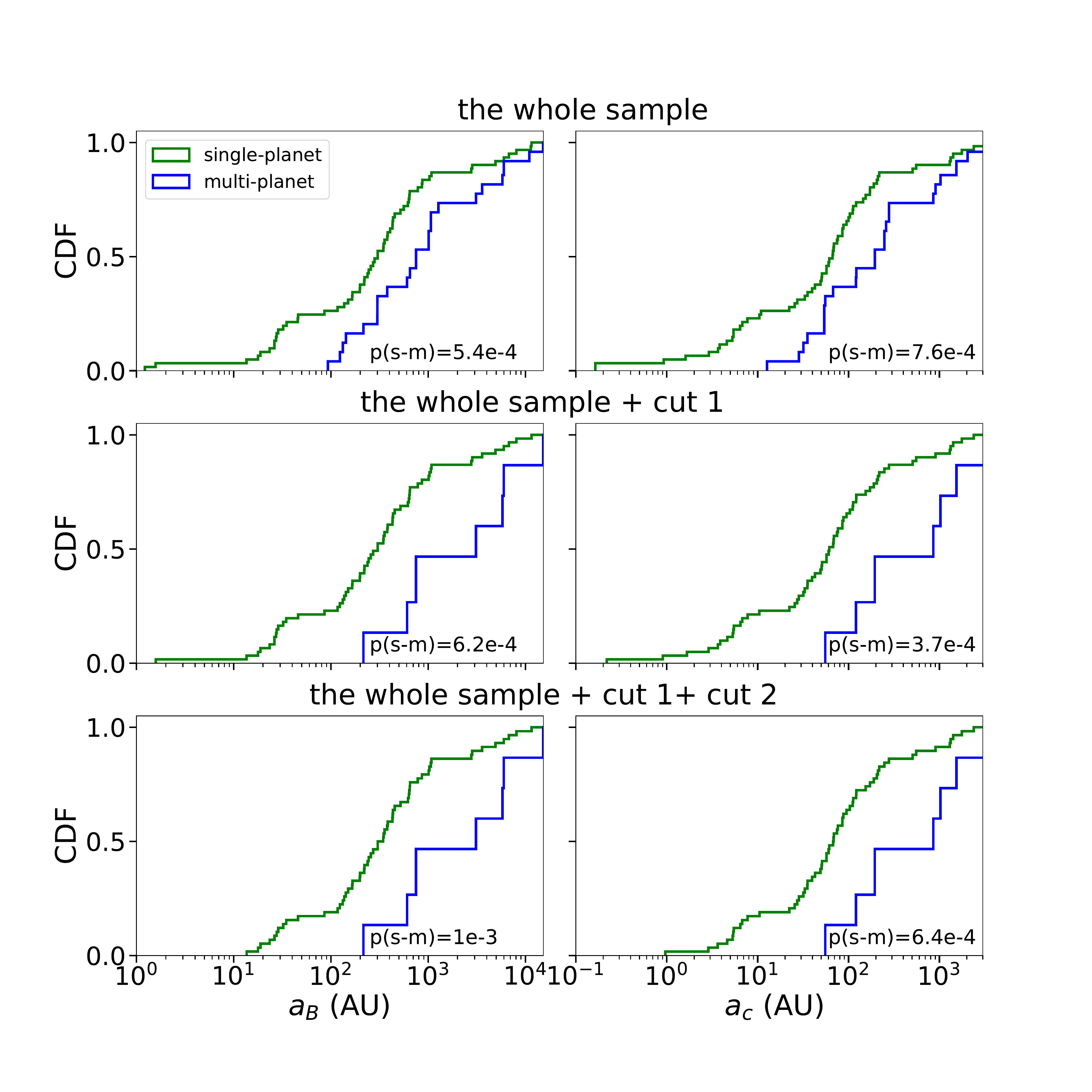}
\end{center}
\caption{\Su{Cumulative Distribution Functions (CDF) of the semi-major axis of binary orbit ($a_B$, first column), and the critical distance for long-term orbital stability ($a_c$, second column). {\xxx \emph{Top row}: ``raw'' sample defined without taking into account potential observational biases. \emph{Middle row}: removing planets below the median detection limit of the singles ``cut-1"). \emph{Bottom row} with both cut-1 and cut-2, i.e., removing systems with RV rms$>$100 ms$^{-1}$. Note that, in the middle and bottom rows, the single- and multiple-planet samples are defined according to the number of planets left \emph{after} applying the different detection limit cuts (so that, for example, some ``multiple systems" of the top row are categorized as ``single-planet" systems in the middle row).}
{\xx In each panel, we perform the two sample KS test and print the p value of the test between singles and multiples, i.e., p(s-m).}
}
\label{fig:s_m_star}}
\end{figure*}

\begin{figure*}
\begin{center}
\includegraphics[width=1\textwidth]{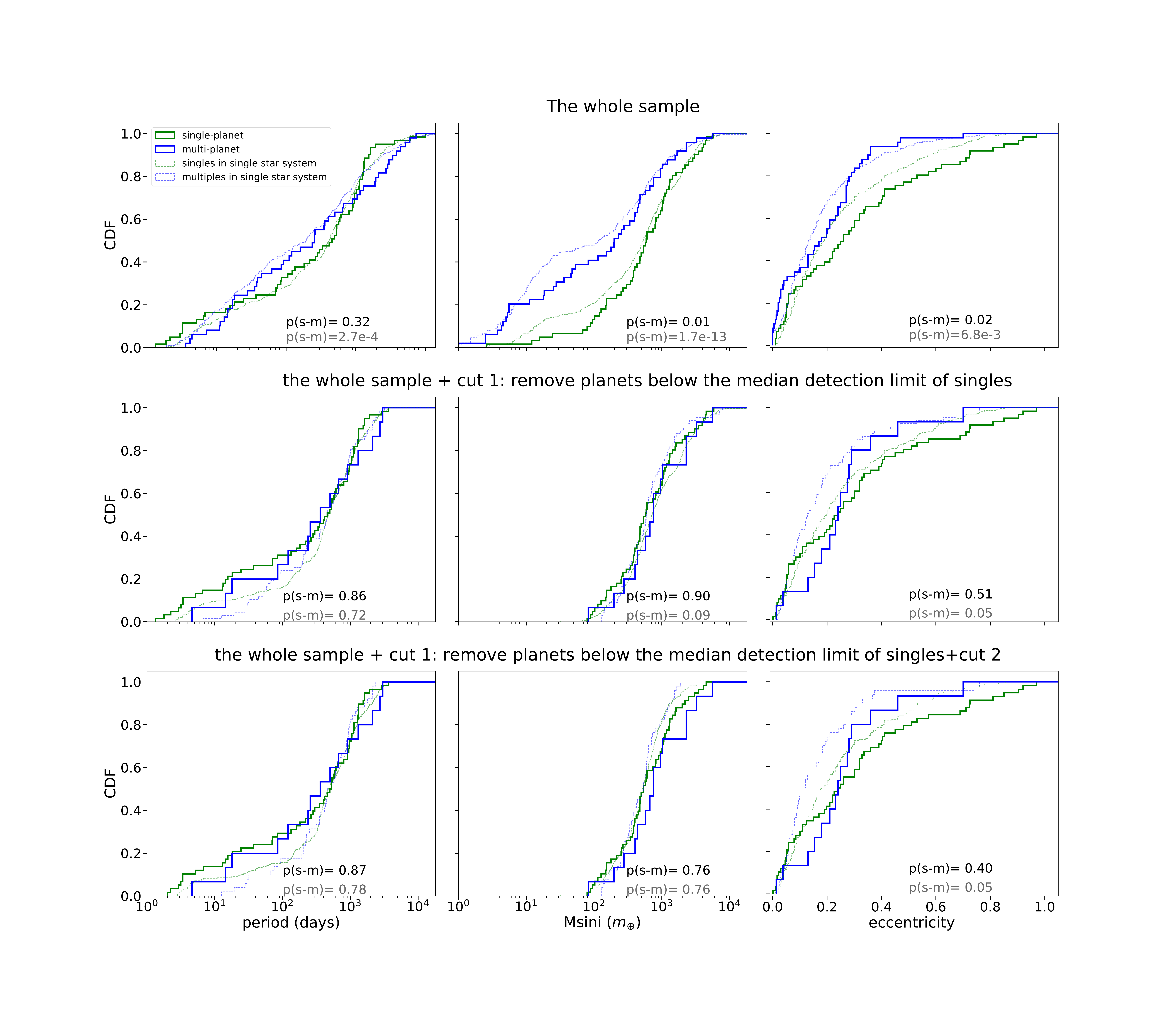}
\end{center}
\caption{\xie{Cumulative Distribution Functions (CDF) of planetary orbital periods ( first column), mimumum masses (Msini, second column) and orbital eccentricities (third column) for single (green) and multiple (blue) planet systems. {\xxx \emph{Top row}: ``raw'' sample defined without taking into account potential observational biases. \emph{Middle row}: removing planets below the median detection limit of the singles ``cut-1"). \emph{Bottom row} with both cut-1 and cut-2, i.e., removing systems with RV rms$>$100 ms$^{-1}$. Note that, in the middle and bottom rows, the single- and multiple-planet samples are defined according to the number of planets left \emph{after} applying the different detection limit cuts (so that, for example, some ``multiple systems" of the top row are categorized as ``single-planet" systems in the middle row).}
{\xx In each panel, we perform the two sample KS test and print the p value of the test between singles and multiples, i.e., p(s-m).
For comparison, in the three right columns, we also plot  the results of RV planets in single star systems (light curves) from the NASA Exoplanet Archive and print the corresponding p values (grey numbers)}.}
\label{fig:s_m_planet}}
\end{figure*}

\section{Results} \label{sec:results}

\subsection{Single Planets  VS. Multiple Planets} \label{subsec:singles}

{\xie 
We first compare the properties of single and multiple planet systems.
Figure \ref{fig:s_m_pmp} shows their period-mass diagrams.
{\xxx In order to visualize the aforementioned potential RV detection bias, } we also overplot the detection limits of each individual system, i.e., the boundaries between the recovery and no-recovery regions in the RV signal injection-recovery test (e.g., Figure \ref{fig:figa0}).
We then estimate the medians of the detection limits {\xxx for each subsample of single-planet and multiple-planet systems (green and blue curves in Figure \ref{fig:s_m_pmp}), which are defined by the 50\% completeness contours, meaning that} planets below them in the period-mass diagram could not be detected in half systems of the samples.
To reduce the effects of such observational biases, we remove planets in the regions with low completeness.
Specifically, we implement a cut (hereafter labelled ``cut-1") for which we remove planets below the median detection limit of single-planet systems (the green curve) both in the single {\emph and} the multiple subsamples.
{\jw Note that the resulting planetary parameters actually depend on the number of planets considered when fitting the RV data (e.g., \citet{2013ApJS..208....2W}
). 
For simplicity, we ignored this effect in this paper and adopted the same planetary parameters before and after ``cut-1".
}
In addition, we also consider another cut (hereafter ``cut-2") for which we remove systems with very poor RV data (i.e., rms $>100$ ms$^{-1}$), since the detection limits of such poor RV systems deviate too much from the median, and the detectable areas are too small in the period-mass diagram (e.g. system HD 87646, see Fig. \ref{fig:figa0}).
{\xxx Figure \ref{fig:s_m_pmp} shows that there are 61 single-planet systems, of which 53 are above the detection limit of cut-1 (the green curve), while there are 53 planets in multiple systems, of which only 26 lie above cut-1.}
{\xx Note that singles and multiples in Figure \ref{fig:s_m_pmp} are initially defined according to the number of planets actually observed in the systems.} {\xxx However, because of the significant fraction of planets that do not pass the cut-1 test, when analysing statistical properties, we also consider another classification, for which we define single- and multiple-planet systems according to the number of ``remaining" planets after applying this cut-1 criterion (see section 3.1.2). With this alternative classification, some initial multiple-planet systems could be “converted” into single-planet ones (see Fig.\ref{fig:s_m_star} {\Su and Fig.\ref{fig:s_m_planet}}).
}}

{\xie
\subsubsection{multiple-planet systems prefer wide binaries}
Previous studies have hinted at that multiple planet systems are preferentially found in wide binaries \citep{2012A&A...542A..92R}. 
{\xxx Here, we revisit this trend with our larger and updated sample.
The three panels in the first column} of Figure \ref{fig:s_m_star} compare the cumulative distribution functions (CDF) of binary separations ($a_B$) for single- and multiple-planet systems. 
{\xxx Already in the raw un-debiased sample there is clear tendency for multiple-planet systems to have larger $a_B$ values as compared to that of single-planet systems. 
{\jw This trend is pronounced  with a KS test p value of {\Su $= 0.00054$}, and it is still significant (p value of {\Su$= 0.00062$}) when attempting to debias the sample by applying cut-1, or with both cut-1 and cut-2 (p value of {\Su$= 0.001$}). } Our results thus clearly confirm that multiple-planet systems are more prevalent in wide binaries whereas single-planet ones are, on average, found in tighter binaries.} 

{\jw In addition, we also consider the dynamical stability limit, $a_c$, defined as the critical distance to the primary star beyond which orbits become unstable due to the perturbations of the companion star. 
We estimate $a_c$ using the empirical formula given by \citet{1999AJ....117..621H}, which is a function of binary mass ratio, orbital semi-major axis ($a_B$) and eccentricity ($e_B$).
We adopted a typical $e_B=0.3$ \citep{1991A&A...248..485D} if the binary eccentricity is not available from the source catalog.
The right panels of Figure \ref{fig:s_m_star} compare the $a_c$ distributions between single and multiples.
We see that multiple planetary systems not only prefer "physically wide" (large $a_B$) binaries but also "dynamically wide" (large $a_c$) binaries.
This trend is significant with a KS test p value of less than $0.001$ regardless of whether cut-1 and cut-2 are applied or not.}
}

{\xie
\subsubsection{period, mass and eccentricity distributions}
The three {\xxx columns} of Figure \ref{fig:s_m_planet} compare the distributions of planet properties, i.e., orbital period, minimum mass and eccentricity for single- and multiple-planet systems.
As can be seen, singles and multiples show similar period distributions regardless of whether the cuts {\xxx correcting for observational bias are applied or not}.
For mass and eccentricity, singles, as compared to multiples, seem to be more massive (p value $= 0.01$) with larger eccentricities (p value {\Su $= 0.02$})  with a statistical confidence level about 2-sigma if using the whole ``raw" sample.
{\xxx However, this trend disappears when correcting for observational bias by applying cut-1 and cut-2, for which the mass and eccentricity distributions become statistically indistinguishable (KS test p values larger than p value of {\Su 0.4}) between singles and multiples. As a consequence, the initial trend in the raw data is probably only due to the fact that an important fraction of the smaller planets with lower eccentricities observed in multiple-systems could not have been detected in the singles (see Figure \ref{fig:s_m_pmp}).}

{\xxx Note that, as mentioned earlier, in the second and third rows of Figure \ref{fig:s_m_planet}, systems are classified as singles/multiples according to the  number of planets remaining after correcting for detection limits (cut-1 and cut-2). More specifically, 11 multiple-planet systems from the raw sample are ``converted" into singles after applying cut-1 ({\Su 61} singles and 15 multiples in the second row) and two more if further applying cut-2 ({\Su 58} singles and {\Su 15} multiples in the third row).
Nevertheless, we find similar results regarding the statistics of planetary period, mass and eccentricity distributions (no difference between single and multiple planet systems) even when keeping the initial criterion for defining the two subsamples (i.e., a ``multiple" system is still labelled as such even it only has one planet surviving cuts 1 and 2), which confirms the robustness of this result.
}
In summary, for giant planets in binary star systems (all planets that lie above the median detection limit have Msini $> 60 m_\oplus$), we find that their 1-D distributions in period, mass and eccentricity is similar in single and multiple-planet systems. 
As for smaller planets (Msini $< 60 m_\oplus$), the conclusion cannot be drawn with current RV data.
}

{\xie
\subsubsection{a gap in the period-mass diagram of single-planet systems? }

Although singles and multiples do not show significant differences in the 1-D distribution of either period or mass (Fig.\ref{fig:s_m_planet}), they seem to differ significantly in the 2-D distribution of the period-mass diagram. 
As can be seen in Figure \ref{fig:s_m_pmp}, there are {\Su $N_s=53$}  singles above the median detection limit (green curve) but none is located in the middle shaded rectangle region (period = 10-300 days and Msini = 160-640 $m_\oplus \sim$ 0.5-2.0  $m_{\rm Jupiter}$).
For comparison, there are {\Su$N_m=23$} multiples above the green detection limit curve and 5, a fraction of {\Su$f=0.22$ }are located in the middle shaded rectangle. 
If we adopt a null hypothesis by assuming that singles and multiples follow the same 2-D distribution in the period-mass diagram, then we can use binominal distribution to access the significance of the observed gap.
The probability that no more than $m$ points are found in the gap follows
\begin{equation}
    P_{gap}=\sum_{k = 0}^{m}C_{n}^{k}(1- p)^{(n-k)}p^k,
    \label{pgap}
\end{equation}
where $n$ is the total points, $m$ is the number of points observed in the gap, $C_{n}^{k}$ the combination coefficient and $p$ the nominal probability each point would be located in the gap under the null hypothesis.
Through out this paper, we adopt {\Su$p = f=0.22$}.
Here, there are {\Su$n=53$} singles being considered, then the probability that none ($m=0$) is found in the gap by chance is {\Su $P_{gap}=(1-p)^{53}=2.28 \times 10^{-6}$}.

Nevertheless, above simple calculation implicitly assumes that planets in the rectangle are fully detectable (100\% detection completeness) for all single systems, which is not the case in reality.
By taking into account the effects of detection completeness of individual systems, the probability that no more than $m$ out of $n$ singles are found in the gap by chance is modified as (marked with a superscript `*')
\begin{equation}
    P_{gap}^*=\sum_{k = 0}^{m}\sum_{l = 1}^{C_{n}^{k}}\left(\prod_{i = 1}^{n-k}(1-p\times D_{l,i})\prod_{j = 1}^{k}p\times D_{l,j}\right),
    \label{pgap*}
\end{equation}
where $D_{l,i}$ and $D_{l,j}$ are the coefficients to correct the detectability of individual systems in the $l$-th case in $C_{n}^{k}$ combinations.

If we assume that planets, if any, are evenly distributed in the rectangle, then in practice, $D_{l,i}=A_i/A_{tot}$, where $A_{tot}$ is the total area 
of the rectangle and $A_i$ the area of the part of rectangle that is above the detection limit of individual systems.
For example (see Fig. \ref{fig:figa0}), HD 41004 B has a poor RV precision (rms=589.2 ms$^{-1}$), and thus zero detectabilitiy in the rectangle, i.e., $D_{l,i}=A_i/A_{tot}=0$; 
in contrast, HD 41004 A has a good RV precision (rms=9.69 ms$^{-1}$) and 100\% detectabilitiy in the rectangle, i.e., $D_{l,i}=A_i/A_{tot}=1$.
Using the above Equation \ref{pgap}, the probability that none of singles is found in the gap by chance is {\Su$P_{gap}^*=1.15 \times 10^{-5}$}, which is over {\Su five } times of the p value ({\Su$P_{gap}=2.28 \times 10^{-6}$}) if observational biases of individual systems were ignored as mentioned before.

{\xx Note, as mentioned before, that there are two criteria for classifying systems as singles and multiples. 
One is based on the number of planets \emph{actually observed} in the system.
The other one is based on the number of remaining planets after applying detection limit cuts.
{\xxx Contrary to the analysis of the distributions of binary separation or planetary orbital parameters (Figure \ref{fig:s_m_star} and \ref{fig:s_m_planet}), we here chose the first criteria for defining the two subsamples of single- and multiple-planet systems.
The main reason is that the second criterion is statistically too drastic}, converting too many multiples ({\Su 8 out of 23}) into singles, resulting in (1) possible contamination of any potential feature (e.g., the gap) that is intrinsic to singles, and (2) too few remaining multiples for statistical study in the  2-D period-mass diagram. 

Granted, the first criterion also has its own weakness, mainly that some of the observed singles could be misidentified because of potential undetected planets. Nevertheless, the effect of such an observational bias can be relatively straightforward to quantify.
{\xxx Assuming that $N_{S-B}$ observed singles in the left panel of Figure \ref{fig:s_m_pmp} are in reality ``hidden" multiples because of the current detection efficiency of multiples just removes $N_{S-B}$ planets around the gap without adding any planet to it in the left panel of Fig. \ref{fig:s_m_pmp},} the gap itself should basically remain as long as $N_{S-B}$ is not particularly large compared to the total number of singles ({\Su$N_S=61$}). 

We obtain a rough estimate of $N_{S-B}$ by injecting all the second strongest planet signals from the multiples (right panel of {\xxx Fig. \ref{fig:s_m_pmp}} ) into the RV data of each single (left panel of {\xxx Fig. \ref{fig:s_m_pmp}} ).
We find that, on average, a fraction $\sim43\%$ of the injected planet signals could be recovered, indicating that $\sim43\%*N_S = 26$ of the observed singles are likely to be real singles.
For the remaining $\sim57\%*N_S = 35$ uncertain ones, if assuming a single/multiple ratio comparable to that for the certain ones (i.e, {\Su$\sim43\%*N_S/N_M = 26/49$}), then 23 of them should be intrinsic multiples, namely $N_{S-B}=23$.
We then estimate how the significance of the gap would be modified.
Given $N_{S-B}=23$ observed singles are re-classified as multiples, the fraction of multiples in the gap reduces to {\jw $f=5/(23+23)=0.10$}, and the number of singles reduces to 38. 
Putting these two reduced numbers into Equations \ref{pgap} and \ref{pgap*}, we obtain $P_{gap}=0.017$ and $P_{gap}^*=0.028$ (about 2 sigma confidence).
Recalling the previous  {\Su$P_{gap}=2.28\times10^{-6}$ and $P_{gap}^*=1.15\times10^{-5}$}, we therefore conclude that the effect of mixing undetected multiples in observed singles is unlikely to destroy the gap in singles though significantly reduce the statistical confidence.
}
}

\begin{figure}
\plotone{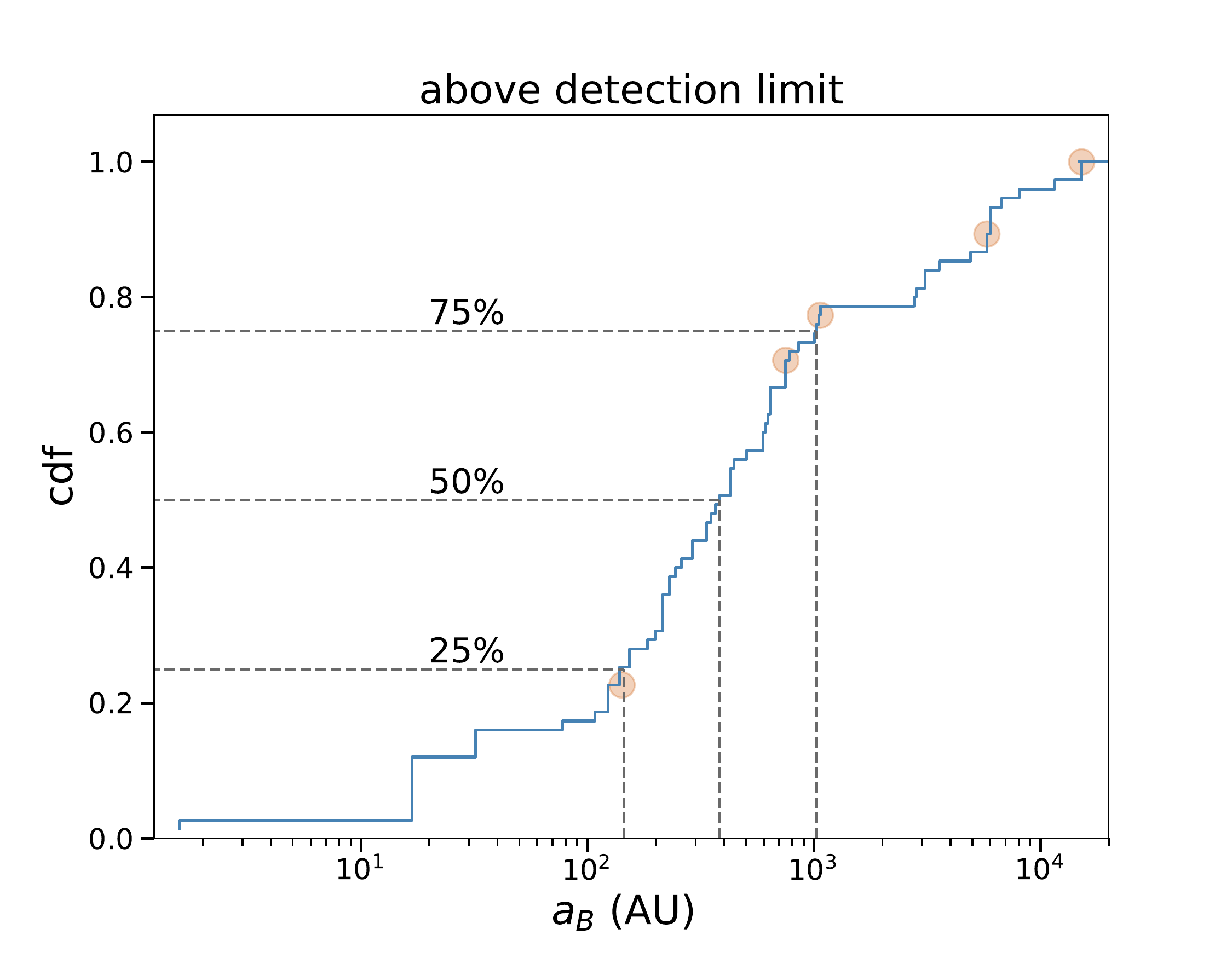}
\caption{{\xie Cumulative distribution function (CDF) of the orbital semi-major axis $a_B$ of the binaries with planets lying above the detection limit as defined by the green curve in Figure \ref{fig:s_m_pmp}.
Binaries with planets located in the period-mass gap (the shaded rectangle in Figure \ref{fig:s_m_pmp}) are marked with orange circles. 
For comparison, the 25, 50 and 75 percentages ({\Su $a_B^{25\%}=155$ AU, $a_B^{50\%}=383$ AU,  and $a_B^{75\%}=1020$ AU}) are also marked. 
}
\label{fig:cdf_aB}}
\end{figure}

\begin{figure*}
\begin{center}
\includegraphics[width=1\textwidth]{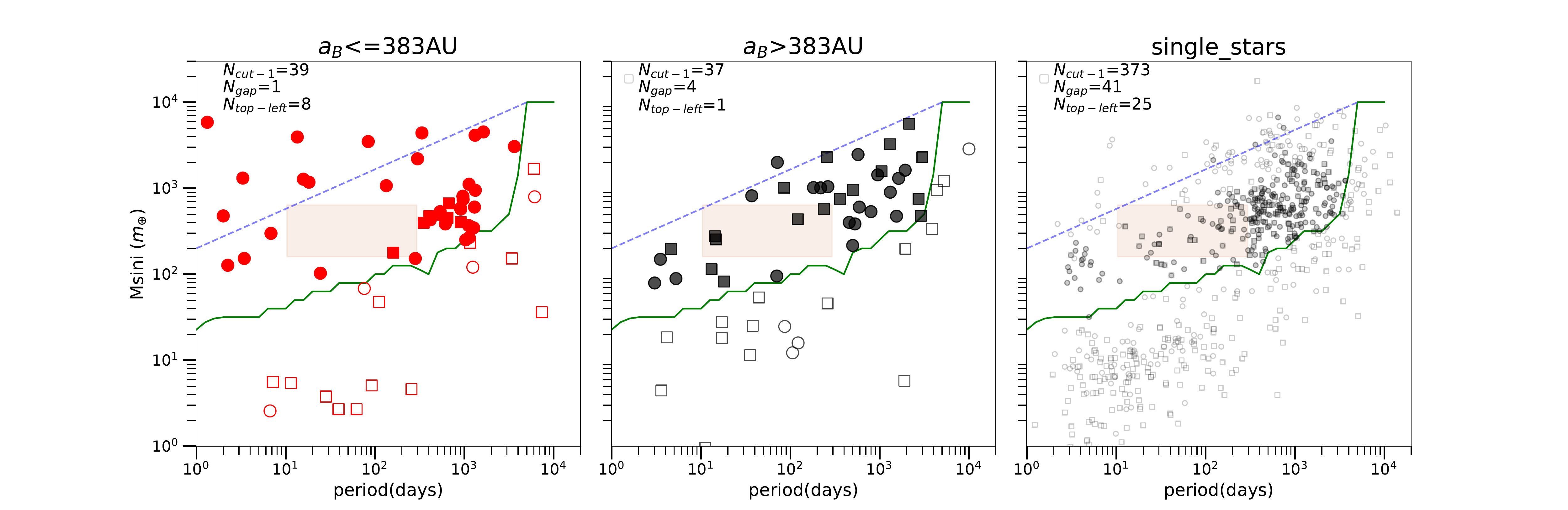}
\end{center}
\caption{{\xie 
Mass (Msini)-period diagrams of RV planets in close binaries (left panel), wide binaries (middle panel) and single star systems (right panel). 
Here, close/wide binaries are defined as $a_B$ smaller/larger than the 50 percentage, {\Su$a_{B}^{50\%}=383$ }AU, of the sample. 
In each panel, the green curve is the detection limit as defined by the same green curve in Figure  \ref{fig:s_m_pmp}, the blue dashed line is an empirical fit to the upper envelope of the planet distribution in the wide sub-sample (panel), and the shaded rectangle markers the same planet desert as identified in Figure \ref{fig:s_m_pmp}. 
Planets above/below the green curve are marked with filled/open circles (single planet systems) and squares (multiple planet systems), respectively.
{\xx In the top-left of each panel, we print the number of planets above the detection limit ($N_{cut-1}$), the number of planets in the gap ($N_{gap}$), and the number of planets on the top-left side of the blue dashed line ($N_{top-left}$)}. 
}
\label{fig:cw381}}
\end{figure*}

\begin{figure*}
\begin{center}
\includegraphics[width=0.99\textwidth]{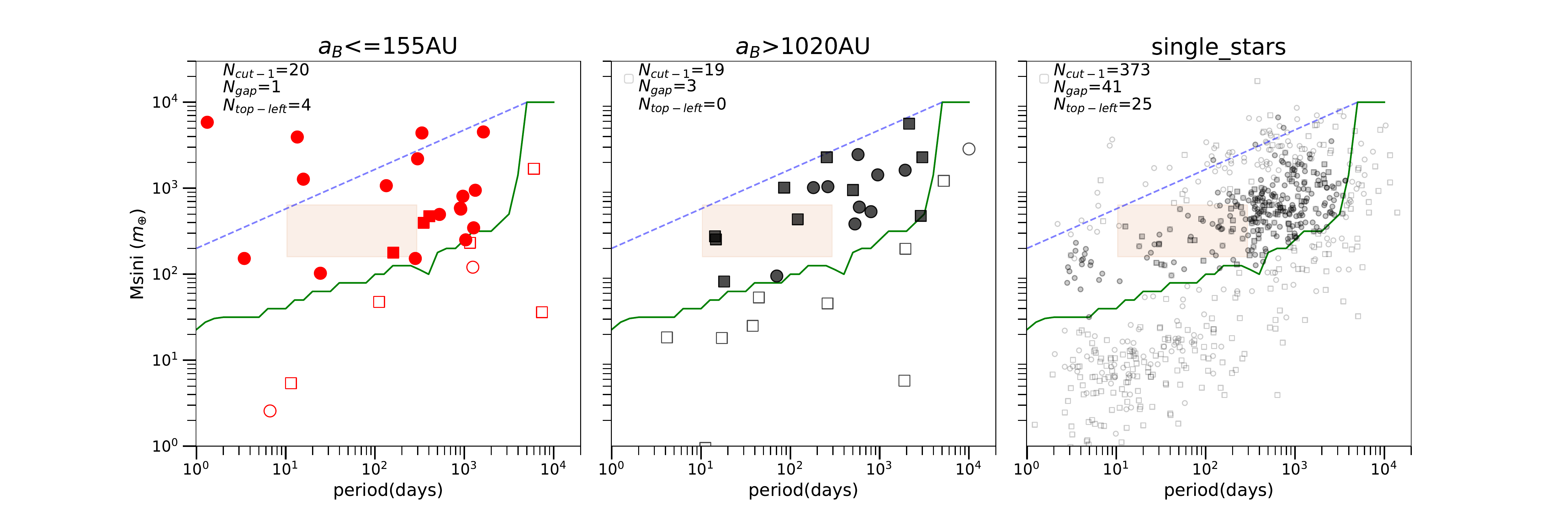}
\end{center}
\caption{ {\xie Similar to Figure \ref{fig:cw381} but here, close/wide binaries are defined as $a_B$ smaller/larger than the 25/75 percentage, {\Su$a_{B}^{25\%}=155$ AU/$a_{B}^{75\%}=1020$}AU, of the sample.
}
\label{fig:cw145}}
\end{figure*}

\subsection{Close VS. Wide Binaries}

{\xie
\subsubsection{Is the gap more prominent in close binaries?}

{\xxx As there seems to be a link between planetary multiplicity and binary star separation (section 3.1.1), we here investigate whether the period-mass gap found in single-planet systems} (section 3.1.3) is also dependent on binary separation. 
As before, we only consider systems with ``good detectability", i.e., systems that lie above the median detection limit of singles (the green line in Figure \ref{fig:s_m_pmp}). {\xxx Even though the gap itself lies in its entirety above the green line, this criterion will change the number of planets outside of the gap that will serve to estimate the statistical significance of the gap.}
Figure \ref{fig:cdf_aB} shows the cumulative distribution of the binary separations ($a_B$) of these systems.
There are five binaries which host planets in the gap, and their separations ({\Su$a_B=$ 142, 750 ,1065, 5796 and 15184 AU}) are marked in Figure \ref{fig:cdf_aB}. 
Clearly, these gap planets tend to reside in binaries with relatively wide separations, {\xxx meaning that the gap is preferentially found in close binaries.
To further demonstrate this point, we conducted two additional analyses as follows.

In the first analysis, we divided the whole sample into two ``close" and ``wide" binary sub-samples, taking as a boundary the median binary separation for our whole sample $a_{B}^{50\%}$={\Su383} AU.} 
 Figure \ref{fig:cw381} shows the period-mass diagrams of planets in these two subsamples.
{\xxx As can be seen, in the wide subsample, out of the {\Su37} planets above the detection limit line, there are 4 lying in the gap, corresponding to a fraction of {\jw $\sim11^{+8}_{-5}\%$} (errorbars reflecting Poisson uncertainties).}
In contrast, in the close-binaries subsample, there is only 1 planet out of {\Su39 ($\sim2^{+6}_{-2}\%$) }that lies in the gap, and actually near the lower boundary of it.

In the second analysis, {\xxx we consider a more stringent definition of our} close and wide subsamples: a binary is in the close subsample if its $a_B$ is smaller than the 25\% percentage of the sample ({\Su$a_{B}^{25\%}=155$} AU), and is in the wide subsample if its $a_B$ is larger than the 75\% percentage of the sample ({\Su$a_{B}^{75\%}=1020$ }AU).
{\xx As can be seen in Figure \ref{fig:cw145}, there are now 3 out of {\Su19 ($\sim15^{+15}_{-8}\%$)} planets in wide binaries that reside in the gap.
In contrast, the number of gap planets is one ({\Su1 out of 20 or $\sim5^{+11}_{-4}\%$} if considering Poisson uncertainty) for the close-binary subsample.}

{\xxx Both analyses show that the period-mass gap identified for single-planet systems is also preferentially found in close-binaries, even though there is a relatively large Poisson uncertainty given the relatively small size of the subsamples}.

}

\begin{figure*}
\begin{center}
\includegraphics[width=0.99\textwidth]{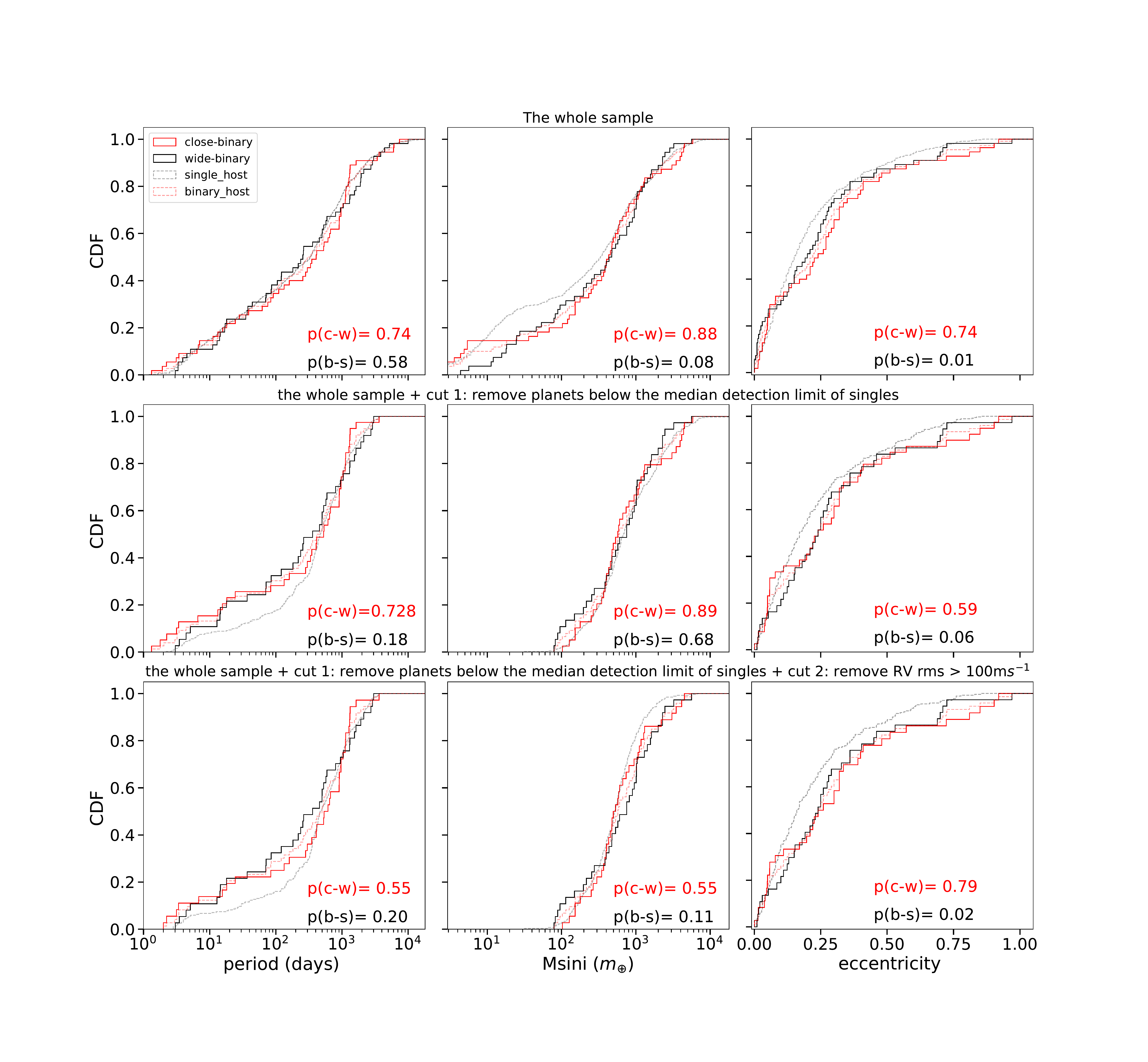}
\end{center}
\caption{ {\xie Comparison of planetary orbital periods (first column), mimimum masses (Msini, second column) and orbital eccentricities (third column) between close (red) and wide (grey) binaries in samples without any observational bias cut (top row), with cut-1, i.e., removing planets below the median detection limit of the singles (middle row) and with both cut-1 and cut-2, i.e., removing systems of RV rms$>$100 ms$^{-1}$ additionally (bottom row). 
In each panel, we perform the two sample KS test and print the corresponding p value.
Here, close/wide binaries are defined as in Figure \ref{fig:cw381}, namely, $a_B$ smaller/larger than the 50 percentage, {\Su$a_{B}^{50\%}=383$} AU, of the sample.
}
\label{fig:cw381cdf}}
\end{figure*}

\begin{figure*}
\begin{center}
\includegraphics[width=0.99\textwidth]{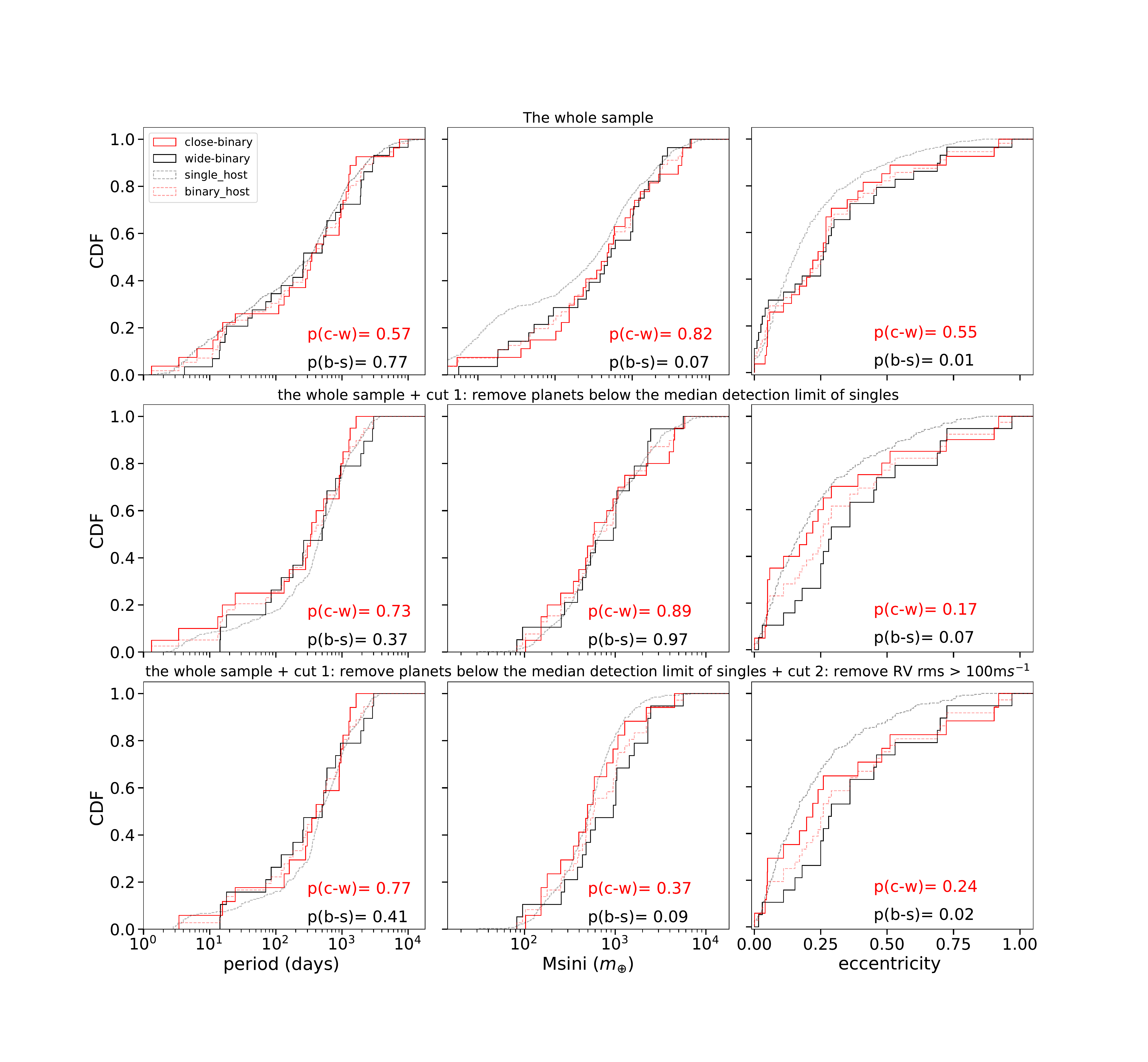}
\end{center}
\caption{{\xie Similar to Figure \ref{fig:cw381cdf} but here, close/wide binaries are defined as in Figure \ref{fig:cw145}, namely, $a_B$ smaller/larger than the 25/75 percentage, {\Su$a_{B}^{25\%}=155$ AU/$a_{B}^{75\%}=1020$} AU, of the sample.
}
\label{fig:cw145cdf}}
\end{figure*}

{\xie
\subsubsection{period, mass and eccentricity distributions}

{\xxx As can be seen in Figures \ref{fig:cw381cdf} and \ref{fig:cw145cdf}, there is no statistically significant difference in the 1-D distribution of planetary period, mass and eccentricity between systems in close and wide binaries. 
This result holds for the two different definitions of the close and wide subsamples and regardless whether observational bias cuts are taken into account or not.
These results also clearly show up when considering} the p values of the KS test between close and wide binaries, which are all greater than 0.5 in Figure \ref{fig:cw381cdf} and greater than {\Su 0.1} in Figure \ref{fig:cw145cdf}.

{\xxx There is, however, a difference between the 2 subsamples for the 2-D distribution of the planetary period-mass diagram, {\xx which is different from} the already identified gap that prefers close binaries over wide binaries (section 3.1.3).
As can be seen in Figures \ref{fig:cw381} and \ref{fig:cw145}, there is a rising upper envelope planets in the period-mass diagram for wide binaries {\xx which is not} found in close-binary subsample. 
This upper envelope} empirically follows 
\begin{equation}
    \rm log_{10}\left(\frac{Msini}{m\oplus}\right) = 0.5\times log_{10}\left(\frac{period}{day}\right) + 2.3 .
    \label{bound}
\end{equation}

Specifically, as shown in Figure \ref{fig:cw381}, only 1 out of {\Su37 ($\sim2^{+6}_{-2}\%$)} planets (with filled symbols) is above the blue dashed line (an empirical upper envelope) in the wide subsample, while this fraction number is {\Su8/39 ($\sim21^{+10}_{-7}\%$)} in the close subsample.
Such a difference is still significant ({\Su$0/19\sim0^{+9}_{-0}\%$ of wide v.s. $4/20\sim20^{+16}_{-10}\%$ of close}) in Figure \ref{fig:cw145}, in which close/wide binaries are defined with a more stringent criterion.
Note that this result that massive short period planets are rare in wide binaries cannot be explained by observational bias because these planets are the easiest ones to be detected. The top-left region of the period-mass diagram has indeed the highest detection efficiency as shown in Figure \ref{fig:s_m_pmp}.
}

{\xx 
\subsection{Single Stars VS. Binary Stars}

{\xxx For the sake of comparison, we also performed the above analyses to RV planets in single star systems, which were retrieved from the https://exoplanetarchive.ipac.caltech.edu archive.}

{\xx As can be seen in the three right columns of Figure \ref{fig:s_m_planet} (dashed curves), {\xxx the 1-D period and mass distributions of planets become indistinguishable between single and multiple planet systems after applying the debiasing cut-1 and cut-2, a result similar to the one obtained for binaries. Nevertheless, we note that, contrary to binaries, the 1-D distribution in eccentricity remains different between singles and multiples even after debiasing, with a} p value of 0.05 ($\sim$ 2 sigma confidence level).
    
{\xxx The fraction of planets in the period-mass gap is 41/373$\sim11^{+2}_{-2}\%$ for single stars (Figures \ref{fig:cw381} and \ref{fig:cw145}), a value close to the one found for wide binaries ($\sim11^{+8}_{-5}\%$), strengthening} the result found in section 3.2.1 that the gap is preferentially found in close binaries.

{\xxx The rising upper envelope in the period-mass diagram identified for} planets in wide binaries {\xxx is also present for planets around single stars (right panels of Figures \ref{fig:cw381} and \ref{fig:cw145})}.
There are indeed only 25 out of 373 planets, {\xxx corresponding to $\sim6.7^{+1.6}_{-1.3}\%$, above the blue dashed line.  
This fraction is consistent (within 1-sigma) with the one found for} planets in wide binaries ({\Su$1/37\sim2^{+6}_{-2}\%$}), but significantly lower than that of planets in close binaries ({\Su$8/39\sim21^{+10}_{-7}\%$}).
    
As can be seen from the right panels of Figures \ref{fig:cw381cdf} and \ref{fig:cw145cdf}, planets around single stars tend to have lower eccentricities than planets in binaries.
The median planetary eccentricity is 0.16 for single stars, while it is 0.24 in binary star systems, as shown in Figure \ref{fig:cw381cdf} with a KS test p value of {\Su$0.01-0.06$}. 
The result holds when only considering binaries at the two ends of the separation distribution, i.e., $a_B$ smaller/larger than the 25/75 percentage, {\Su$a_{B}^{25\%}=155$ AU/$a_{B}^{75\%}=1020$} AU, of the sample. 
The median eccentricity of planets in binaries increases to {\Su0.26} with KS test p value of {\Su0.01-0.07} as shown in Figure \ref{fig:cw145cdf}. 
This result indicates that stellar binarity increases the eccentricities of exoplanets to a certain extent, and this effect is not confined to close binaries but still significant even in very wide binaries (e.g. {\Su$a_B>1020$ }AU).


}

{\xx
\section{Discussions: Implications to Planet Formation and Evolution}
\label{sec:discussion}

{\xxx Compared to the distribution of planets in wide binaries and single stars,} we find that planets in close ($a_B<$100-300 AU) binaries are preferentially associated with three characteristic} features: a larger fraction of single planets, a rectangle-shaped gap in the period-mass diagram, {\xxx and the presence of massive and short period planets} (Fig. \ref{fig:s_m_planet}, \ref{fig:cdf_aB}, \ref{fig:cw381} and \ref{fig:cw145}).
{\xxx We discuss hereafter the possibility that these 3 features could have the same underlying explanation.}

Close binaries with separation less than a few hundred AU are indeed capable of enhancing planet migration either in the early gas-disk phase \citep{2000IAUS..200P.211K} or in the later disk-free phase e.g., via secular planetary interactions\citep{2007ApJ...669.1298F,2011ApJ...735..109W}, causing planets to {\xxx horizontally}  cross the upper rising envelope in the period-mass diagram identified in Figures \ref{fig:cw381} and \ref{fig:cw145}.
{\xxx In addition, these planets could accrete additional material (e.g., gas, planetesimals or other smaller planets) during their enhanced migration, and could thus grow into much more massive objects \citep{2000IAUS..200P.211K, 2018ApJ...861..116Z}. As a consequence, these planets would populate the initially depleted upper-left part of the period-mass diagram. These enhanced migrations could also lead to an enhanced level of planetary mergers and/or ejections, explaining the lower level of planet multiplicity in close binaries}, as well as their larger eccentricities compared to those in single star systems.
As for the {\xxx rectangle-shaped period-mass} gap, it could be the birth place of those line-crossing planets or planets that were accreted/ejected by them.

As for planets in wide binaries, they share a number of common properties with those around single stars, {\xxx namely a similar fraction (about $10\%$) of planets in the period-mass gap combined to the presence of a similar upper envelope in the same period-mass diagram} (Fig. \ref{fig:cw381} and \ref{fig:cw145}).
The rising upper envelope {\xxx is consistent with} the positive period-mass correlation reported by previous studies \citep{2002ApJ...568L.113Z,2007AJ....134.2061J} {\xxx and with} the predictions of current synthetic models \citep{2004ApJ...604..388I,2009A&A...501.1161M} of planet formation around single stars. 
The common upper envelope between planets in wide binaries and single stars thus may suggest that planet formation and evolution in wide binaries ($a_B>$100-300 AU) {\xxx is mostly} similar to that in single star systems. 
  
Nevertheless, planets in wide binaries differ significantly from planets in single star systems in terms of their eccentricity distribution (Fig. \ref{fig:s_m_planet}, \ref{fig:cw381} and \ref{fig:cw145}). 
{\xxx For single stars}, multiple planet systems have significantly lower eccentricities than single planet systems (the so called ``eccentricity dichotomy"), {\xxx a fact that} has been well established for the Kepler planet sample (dominated by super-Earths and sub-Neptunes, \citep{2016PNAS..11311431X,2019AJ....157...61V,2019AJ....157..198M} and is also seen in RV planets (dominated by Jovian planets) as shown here in Figure \ref{fig:s_m_planet}.
In contrast, {\xxx there is no difference (once the data is debiased) between the eccentricity distributions of single and multiple-planet systems for both our close- and wide binary subsamples. In addition, these eccentricities are, on average, larger} than the eccentricities of planets around single stars (Fig. \ref{fig:s_m_planet}, \ref{fig:cw381cdf} and \ref{fig:cw145cdf}). 
This may imply that there are some additional mechanisms at play to pump up planet eccentricities in wide binaries {\xxx and bring them to the same level as for those in closer binaries.}
{\jw Potential eccentricity drivers could be the Kozai mechanism \citep{1962AJ.....67..591K} and Galactic tides \citep{2013Natur.493..381K}}. 
Future studies, both with larger observational sample and with dedicated numerical models are needed to further unveil the picture of planet formation and evolution in binary star systems.
}

\section{Summary and Conclusion}\label{sec:summary}

{\xie
In this paper, we compile a sample (Table 1) of {\xxx all S-type planetary systems detected by the RV method in order to derive statistical characteristics as a function of key properties of these systems, such as the number of detected planets or binary separation.
Available RV observation data allow us to quantify the planet detection efficiencies of individual systems, enabling us to correct for potential observational bias, notably between the single-planet and multiple-planet sub-samples. We then perform a statistical investigation from two different perspectives.

First, we consider the whole sample of S-type systems and look for statistical properties that depend on planetary multiplicity.
We find that, for the whole S-type sample, single and multiple-planet systems have similar 1-D distributions of both planetary masses and periods. However, they differ significantly in terms of their 2-D period-mass distribution. The main difference is an absence of planets in a rectangle-shaped gap (for planetary periods between 10 and 300 days Msini between 160 and 640 $m_\oplus \sim$ 0.5-2.0  $m_{\rm Jupiter}$) that is present in single-planet systems but not in multiples.}
The gap is statistically significant with a P value of {\Su$P_{gap}=2.28\times10^{-6}$ or  $P^*_{gap}=1.15\times10^{-5}$ }depending if observational bias is ignored (Equation 3) or considered (Equation 4).

{\xxx Second, we look for planet properties that depend on binary separation ($a_B$) and divide our sample into a close  and wide binary subsamples}. 
We find that multiple-planet systems are preferentially found in wide binaries, which is consistent with the previous study by \citep{2012A&A...542A..92R}.
{\xxx We also find that the gap in the planetary period-mass diagram of single-planet systems depends on binary separation: it is more prominent in close binaries with $a_B$ less than a few hundreds AU than in wide binaries or single stars. 
Furthermore, we find that, in the same planetary period-mass distribution, there is a population of massive and short period planets in close-binaries that is not found in wide binaries and single stars, for which the period-mass distribution is limited by a rising upper envelope. Last but not least, we find that the average eccentricity of planets in binaries (both close and wide) is higher than in single stars.

We suggest that enhanced planetary migration, collision and/or ejection in close binaries could explain most of the statistical properties we have identified, but the detailed study of this effect exceeds the scope of the present paper.}

{\xx In summary, our study emphasizes the importance of taking into account individual detection efficiencies when searching for statistical patterns for planets in binaries. 
The results of this paper suggest that binary stars could play a crucial role in shaping the architecture of planetary systems, i.e., reducing planetary multiplicity, exciting planetary eccentricity and modifying the planetary period-mass diagram. 
}}

\begin{acknowledgments}
This work is supported by the National Key R\&D Program of China (No. 2019YFA0405100) and the National Natural Science Foundation of China (NSFC; grant No. 11933001). J.-W.X. also acknowledges the support from the National Youth Talent Support Program and the Distinguish Youth Foundation of Jiangsu Scientific Committee (BK20190005).
\end{acknowledgments}


\appendix

\section{RV Signal Injection-Recovery Tests of the Whole Sample}
\begin{figure*}
\begin{center}
\includegraphics[width=0.85\textwidth]{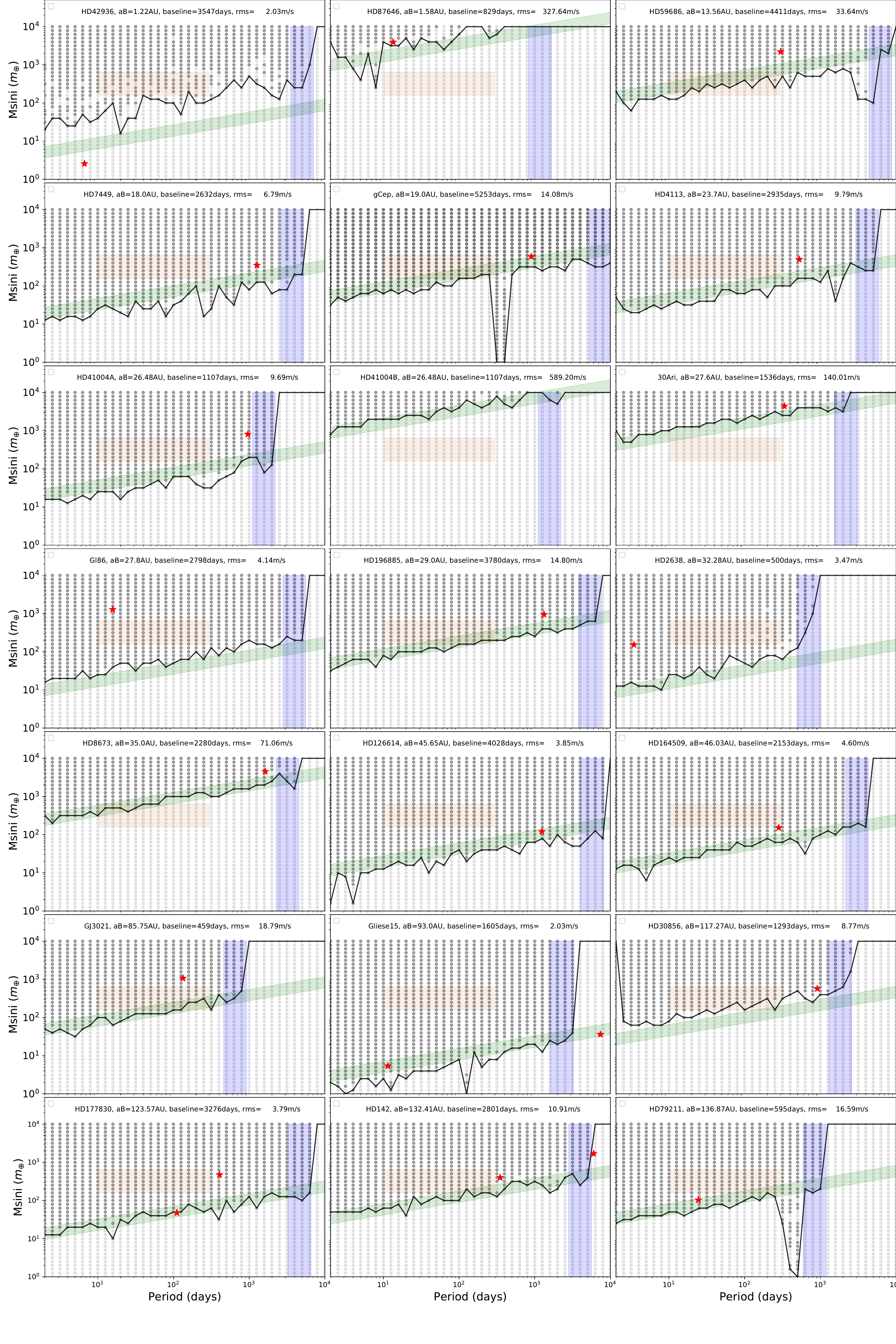}
\end{center}
\caption{ {\jw RV signal injection-recovery tests of 21 binary systems with  {\Su$a_B\leq137$} AU.  
The red star symbols indicate the location of observed planets.
The filled  dark  grey circles (light squares) show  where  the hypothetical  planets  can (can not) be  recovered  from  the  simulated  RVs. 
The green shaded regions mark where $RV_{signal}=(1-2)\times rms$.
The purple shaded regions mark where  $period= (1-2)\times the \ time \ baseline$ of the simulated RV data.
The black curve mark the detect limit of the binary system.}
\label{fig:figa0}}
\end{figure*}

\begin{figure*}
\begin{center}
\includegraphics[width=0.85\textwidth]{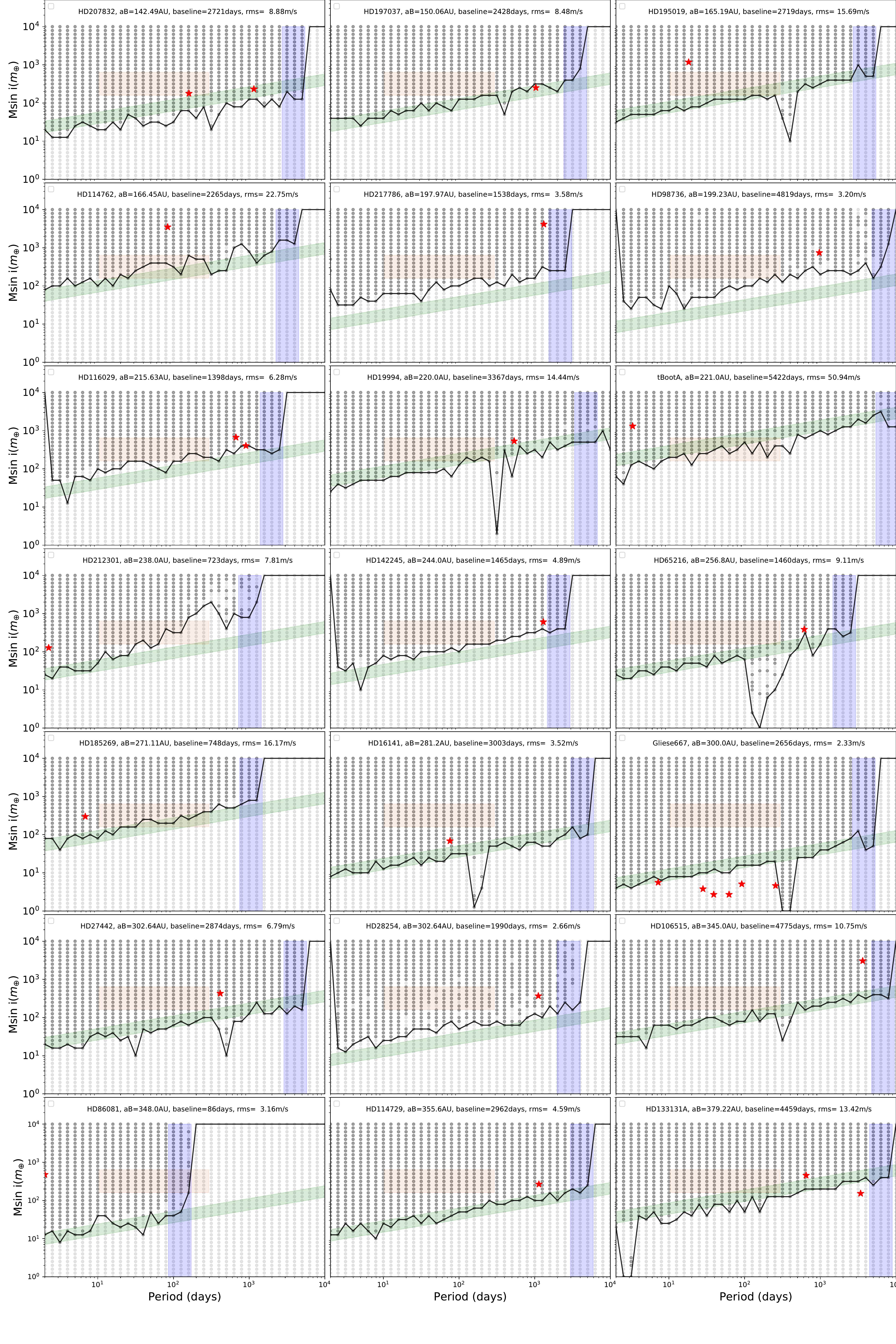}
\end{center}
\caption{{\jw Similar to Figure \ref{fig:figa0} but showing the } RV signal injection-recovery tests of 21 binary systems with {\Su 142 AU $<a_B\leq$ 380 AU. }
\label{fig:figa1}}
\end{figure*}

\begin{figure}
\begin{center}
\includegraphics[width=0.85\textwidth]{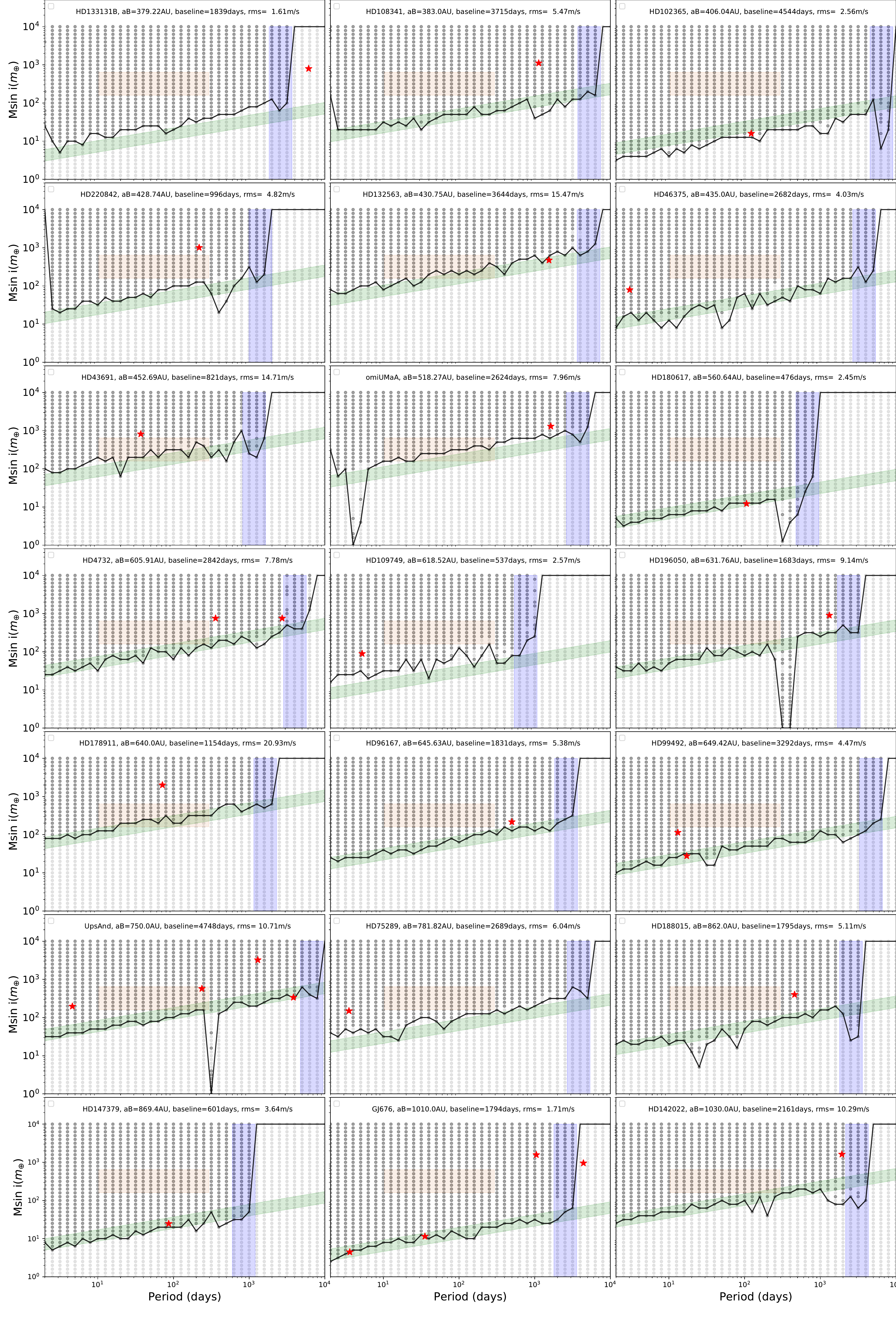}
\end{center}
\caption{{\jw Similar to Figure \ref{fig:figa0} but showing the } RV signal injection-recovery tests of 21 binary systems with {\Su 380 AU $<a_B\leq$ 1030} AU.The caption is  similar to Figure \ref{fig:figa0}.
\label{fig:figa2}}
\end{figure}

\begin{figure*}
\begin{center}
\includegraphics[width=0.85\textwidth]{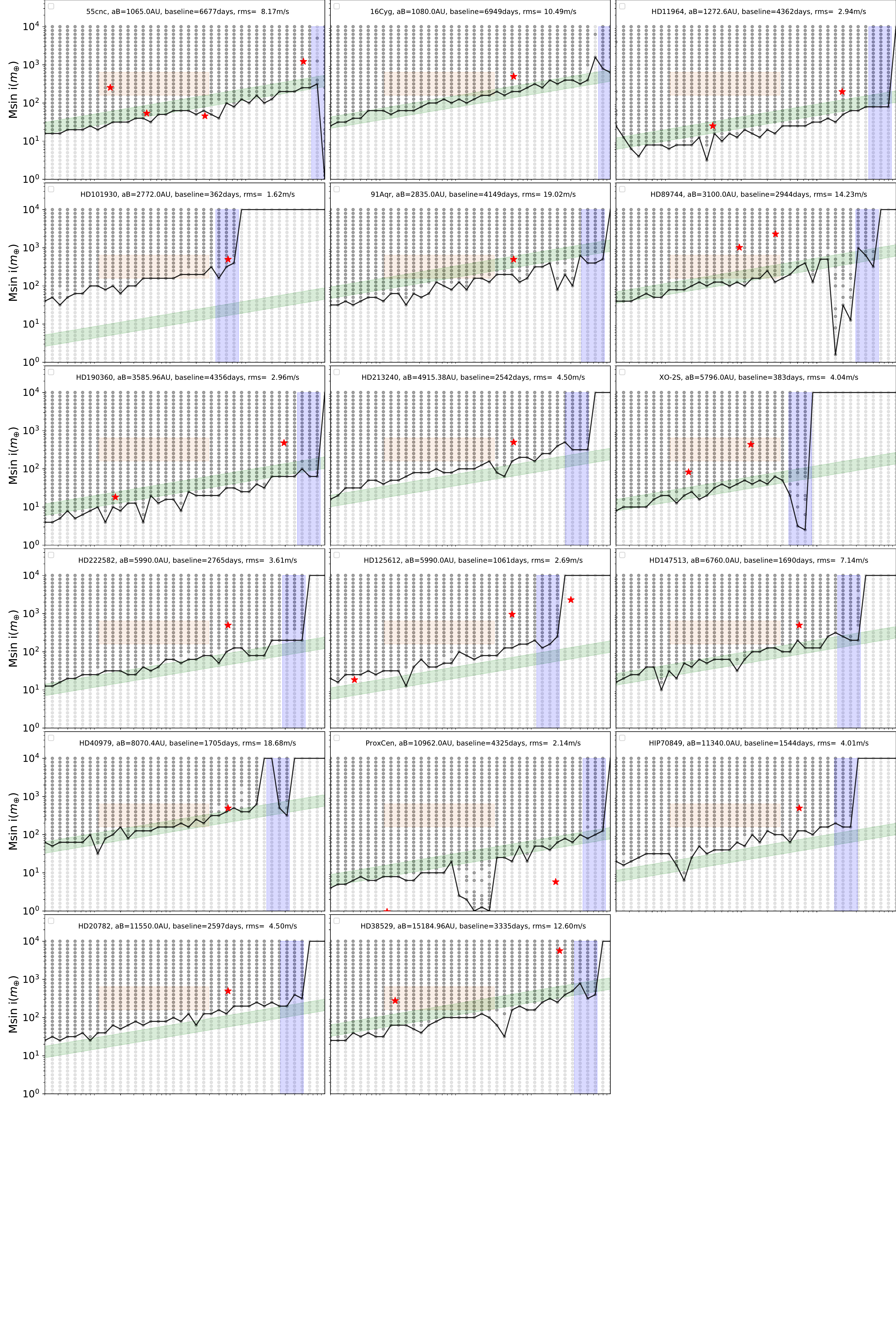}
\end{center}
\caption{{\jw Similar to Figure \ref{fig:figa0} but showing the } RV signal injection-recovery tests of {\Su17} binary systems with $a_B>$ {\Su1030 AU}.The caption is  similar to Figure \ref{fig:figa0}.
\label{fig:figa3}}
\end{figure*}

In Figures \ref{fig:figa0}-\ref{fig:figa3}, we present the results of RV signal injection-recovery tests for all the {\Su80} systems ({\Su110} planets) studied in this paper.

\begin{deluxetable*}{lcccccccccc}
\tablenum{1}
\tablecaption{S type planet in binary stellar systems\label{tab:planet}}
\tablewidth{0pt}
\tablehead{
\colhead{$row\_id$} & \colhead{$Planet\_id$} &  \colhead{$a_{B}$(AU)} & \colhead{$a_{c}$(AU)} &\colhead{$N_p$} &\colhead{ $M_{p}$($M_{J}$)} & \colhead{$a_{p}$(AU)} & \colhead{period(days)} & \colhead{Eccentricity} & \colhead{source} & \colhead{$N-star$} 
}
\decimalcolnumbers
\startdata
1 &  HD 42936 b &  1.22 &  0.2 &  1 &  0.0081 &  0.0662 &  6.6732 &  0.14 &  1 &  2 \\ 
2 &  HD 87646 b &  1.58 &  4.2 &  1 &  12.4 &  0.117 &  13.481 &  0.05 &  1 &  2 \\ 
3 &  HD 59686 A b &  13.56 &  4.1 &  1 &  6.92 &  1.086 &  299.36 &  0.05 &  1 &  2 \\  
4 &  HD 7449 b &  18 &  5.6 &  1 &  1.09 &  2.33 &  1270.5 &  0.92 &  1 &  2 \\  
5 &  gamma Cephei b &  19 &  5.4 &  1 &  1.85 &  2.05 &  903.3 &  0.049 &  1 &  2 \\   
6 &  HD 4113 Ab &  23.7 &  6.7 &  1 &  1.56 &  1.28 &  526.62 &  0.903 &  1 &  2 \\  
7 &  HD 41004 Ab &  26.48 &  5.2 &  1 &  2.54 &  1.64 &  963 &  0.39 &  1 &  2 \\  
8 &  HD 41004 Bb &  26.48 &  3.7 &  1 &  18.4 &  0.0177 &  1.3236 &  0.058 &  1 &  2 \\  
9 &  30 Ari b &  27.6 &  6.3 &  1 &  13.82 &  0.99 &  335.1 &  0.29 &  1 &  4 \\  
10 &  GJ 86 b &  27.8 &  4.4 &  1 &  4.01 &  0.11 &  15.766 &  0.046 &  1 &  2 \\  
11 &  HD 196885 b &  29 &  6.7 &  1 &  2.98 &  2 &  1326 &  0.48 &  1 &  2 \\  
12 &  HD 2638 b &  32.2816 &  6.8 &  1 &  0.48 &  0.044 &  3.438 &  0.041 &  1 &  3 \\  
13 &  HD 8673 b  &  35 &  3.0 &  1 &  14.2 &  3.02 &  1634 &  0.723 &  1 &  2 \\  
14 &  HD 126614 b &  45.6482 &  13.6 &  1 &  0.38 &  2.35 &  1244 &  0.41 &  1 &  3 \\  
15 &  HD 164509 b &  46.0265 &  10.6 &  1 &  0.48 &  0.875 &  282.4 &  0.26 &  1 &  2 \\  
16 &  GJ 3021 b &  85.747 &  22.4 &  1 &  3.37 &  0.49 &  133.71 &  0.511 &  1 &  2 \\ 
17 &  Gliese 15 b &  93 &  15.9 &  2 &  0.017 &  0.072 &  11.44 &  0.27 &  1 &  2 \\  
18 &  Gliese 15 c &  93 &  15.9 &  2 &  0.114 &  5.4 &  7391 &  0.27 &  1 &  2 \\ 
19 &  HD 30856 b &  117.273 &  25.6 &  1 &  1.8 &  2 &  912 &  0.24 &  1 &  2 \\  
20 &  HD 177830 b &  123.57 &  31.5 &  2 &  1.49 &  1.2218 &  406.6 &  0.001 &  1 &  2 \\  
21 &  HD 177830 c &  123.57 &  31.5 &  2 &  0.15 &  0.5137 &  110.9 &  0.35 &  1 &  2 \\  
22 &  HD 142 b &  132.405 &  37.4 &  2 &  1.25 &  1.02 &  349.7 &  0.17 &  1 &  2 \\  
23 &  HD 142 c &  132.405 &  37.4 &  2 &  5.3 &  6.8 &  6005 &  0.21 &  1 &  2 \\  
24 &  HD 79211 &  136.86894 &  27.4 &  1 &  0.3231 &  0.141 &  24.45 &  0.11 &  1 &  2 \\  
25 &  HD 207832 b &  142.493 &  35.2 &  2 &  0.56 &  0.586 &  160.07 &  0.197 &  1 &  4 \\  
26 &  HD 207832 c &  142.493 &  35.2 &  2 &  0.73 &  2.112 &  1155.7 &  0.27 &  1 &  4 \\  
27 &  HD 197037A b &  150.059 &  35.4 &  1 &  0.79 &  2.07 &  1035.7 &  0.22 &  1 &  2 \\  
28 &  HD 195019 b &  165.19 &  37.8 &  1 &  3.7 &  0.1388 &  18.2016 &  0.014 &  1 &  2 \\  
29 &  HD 114762 b &  166.452 &  42.1 &  1 &  10.98 &  0.353 &  83.9151 &  0.3354 &  1 &  2 \\  
30 &  HD 217786 b &  197.97 &  50.4 &  1 &  13 &  2.38 &  1319 &  0.4 &  1 &  2 \\  
31 &  HD 98736 b &  199.23 &  39.7 &  1 &  2.33 &  1.864 &  968.8 &  0.226 &  1 &  2 \\  
32 &  HD 116029 b &  215.63 &  55.4 &  2 &  2.1 &  1.73 &  670.2 &  0.21 &  1 &  2 \\
33 &  HD 116029 c &  215.63 &  55.4 &  2 &  1.27 &  2.144 &  907 &  0.038 &  1 &  2 \\  
34 &  HD 19994 b &  220 &  49.4 &  1 &  1.68 &  1.42 &  535.7 &  0.3 &  1 &  2 \\  
35 &  tau Bootis b &  221 &  5.3 &  1 &  4.13 &  0.046 &  3.3135 &  0.0787 &  1 &  2 \\  
36 &  HD 212301 b &  238 &  69.6 &  1 &  0.4 &  0.036 &  2.2457 &  0.0147 &  1 &  2 \\  
37 &  HD 142245A b &  244 &  72.0 &  1 &  1.9 &  2.77 &  1299 &  0.32 &  1 &  3 \\  
38 &  HD 65216 b &  256.8 &  84.9 &  1 &  1.21 &  1.37 &  613.1 &  0.41 &  1 &  3 \\  
39 &  HD 185269A b &  271.11 &  68.9 &  1 &  0.94 &  0.077 &  6.838 &  0.3 &  1 &  2 \\ 
40 &  HD 16141 b &  281.203 &  66.3 &  1 &  0.215 &  0.35 &  75.82 &  0.28 &  1 &  2 \\
\enddata
\end{deluxetable*}

\begin{deluxetable*}{lcccccccccc}
\tablenum{1}
\tablecaption{continued}
\tablewidth{0pt}
\tablehead{
\colhead{$row\_id$} & \colhead{$Planet\_id$} &  \colhead{$a_{B}$(AU)} & \colhead{$a_{c}$(AU)} &\colhead{$N_p$} &\colhead{ $M_{p}$($M_{J}$)} & \colhead{$a_{p}$(AU)} & \colhead{period(days)} & \colhead{Eccentricity} & \colhead{source} & \colhead{$N-star$} 
}
\decimalcolnumbers
\startdata
41 &  Gliese 667 b &  300 &  52.3 &  6 &  0.0176 &  0.0505 &  7.2 &  0.13 &  1 &  3 \\  
42 &  Gliese 667 c &  300 &  52.3 &  6 &  0.0119 &  0.125 &  28.14 &  0.02 &  1 &  3 \\  
43 &  Gliese 667 d &  300 &  52.3 &  6 &  0.016 &  0.276 &  91.61 &  0.03 &  1 &  3 \\  
44 &  Gliese 667 e &  300 &  52.3 &  6 &  0.0085 &  0.213 &  62.24 &  0.02 &  1 &  3 \\  
45 &  Gliese 667 f &  300 &  52.3 &  6 &  0.0085 &  0.156 &  39.03 &  0.03 &  1 &  3 \\  
46 &  Gliese 667 g &  300 &  52.3 &  6 &  0.0145 &  0.549 &  256.2 &  0.08 &  1 &  3 \\  
47 &  HD 27442 b &  302.64 &  61.2 &  1 &  1.35 &  1.16 &  415.2 &  0.058 &  1 &  2 \\  
48 &  HD 28254A b &  302.64 &  59.8 &  1 &  1.16 &  2.15 &  1116 &  0.81 &  1 &  2 \\ 
49 &  HD 106515 b &  345 &  61.9 &  1 &  9.61 &  4.59 &  3630 &  0.572 &  1 &  2 \\  
50 &  HD 86081A b &  348 &  94.9 &  1 &  1.5 &  0.039 &  1.99809 &  0.0575 &  1 &  2 \\  
51 &  HD 114729 b &  355.602 &  85.5 &  1 &  0.84 &  2.08 &  1135 &  0.32 &  1 &  2 \\  
52 &  HD 133131 Ab &  379.22 &  80.9 &  2 &  1.43 &  1.44 &  649 &  0.32 &  1 &  2 \\  
53 &  HD 133131 Ac &  379.22 &  80.9 &  2 &  0.48 &  4.36 &  3407 &  0.47 &  1 &  2 \\  
54 &  HD 133131 Bb &  379.22 &  79.9 &  1 &  2.5 &  6.415 &  6119 &  0.62 &  1 &  2 \\  
55 &  HD 108341 b &  383 &  95.5 &  1 &  3.5 &  2 &  1129 &  0.85 &  1 &  2 \\  
56 &  HD 102365 b &  406.042 &  99.8 &  1 &  0.05 &  0.46 &  122.1 &  0.34 &  1 &  2 \\ 
57 &  HD 220842A b &  428.739 &  85.4 &  1 &  3.18 &  0.74 &  218.47 &  0.404 &  1 &  2 \\  
58 &  HD 132563 b &  430.75 &  86.7 &  1 &  1.49 &  2.62 &  1544 &  0.22 &  1 &  3 \\  
59 &  HD 46375 b &  435 &  71.5 &  1 &  0.249 &  0.041 &  3.024 &  0.0524 &  1 &  2 \\  
60 &  HD 43691 b &  452.69 &  111.9 &  1 &  2.57 &  0.238 &  36.99913 &  0.085 &  1 &  3 \\  
61 &  omi UMaA b &  518.271 &  103.0 &  1 &  4.1 &  3.9 &  1630 &  0.191 &  1 &  2 \\ 
62 &  HD 180617 b &  560.6406 &  144.5 &  1 &  0.0384 &  0.3357 &  105.9 &  0.16 &  1 &  2 \\ 
63 &  HD 4732 c &  605.9105 &  113.0 &  2 &  2.37 &  4.6 &  2732 &  0.23 &  1 &  2 \\ 
64 &  HD 4732 b &  605.9105 &  120.6 &  2 &  2.37 &  1.19 &  360.2 &  0.13 &  1 &  2 \\ 
65 &  HD 109749 b &  618.5205 &  121.0 &  1 &  0.28 &  0.0635 &  5.24 &  0.01 &  2 &  2 \\  
66 &  HD 196050 b &  631.761 &  156.9 &  1 &  2.83 &  2.47 &  1316.24 &  0.21 &  2 &  3 \\ 
67 &  HD 178911 b &  640 &  173.9 &  1 &  6.29 &  0.32 &  71.49 &  0.124 &  2 &  3 \\  
68 &  HD 96167 b &  645.632 &  172.1 &  1 &  0.68 &  1.3 &  498.9 &  0.71 &  2 &  2 \\
69 &  HD 99492  b &  649.415 &  121.6 &  2 &  0.087 &  0.122 &  17.1668 &  0.13 &  2 &  2  \\  
70 &  HD 99492 c &  649.415 &  121.6 &  2 &  0.359 &  5.4 &  13.1 &  0.1 &  2 &  2 \\  
71 &  Ups And b &  750 &  245.1 &  4 &  0.62 &  0.059 &  4.6171 &  0.013 &  2 &  2 \\  
72 &  Ups And c &  750 &  245.1 &  4 &  1.8 &  0.861 &  237.7 &  0.24 &  2 &  2 \\  
73 &  Ups And d &  750 &  245.1 &  4 &  10.19 &  2.55 &  1302.61 &  0.274 &  2 &  2 \\ 
74 &  Ups And e &  750 &  245.1 &  4 &  1.059 &  5.2456 &  3848.86 &  0.005 &  2 &  2 \\ 
75 &  HD 75289 b &  781.819 &  204.7 &  1 &  0.47 &  0.046 &  3.5098 &  0.021 &  2 &  2 \\  
76 &  HD 188015 b &  862 &  217.6 &  1 &  1.26 &  1.19 &  456.46 &  0.15 &  2 &  2 \\ 
77 &  HD 147379 b &  869.4 &  171.8 &  1 &  0.0777 &  0.3193 &  86.54 &  0.01 &  2 &  2 \\  
78 &  GJ 676 b &  1010 &  247.8 &  4 &  4.95 &  1.8 &  1050.3 &  0.328 &  2 &  2 \\  
79 &  GJ 676 c &  1010 &  247.8 &  4 &  3 &  5.2 &  4400 &  0.2 &  2 &  2 \\  
80 &  GJ 676 d &  1010 &  247.8 &  4 &  0.014 &  0.0413 &  3.6 &  0.15 &  2 &  2 \\ 
\enddata
\end{deluxetable*}

\begin{deluxetable*}{lcccccccccc}
\tablenum{1}
\tablecaption{continued}
\tablewidth{0pt}
\tablehead{
\colhead{$row\_id$} & \colhead{$Planet\_id$} &  \colhead{$a_{B}$(AU)} & \colhead{$a_{c}$(AU)} &\colhead{$N_p$} &\colhead{ $M_{p}$($M_{J}$)} & \colhead{$a_{p}$(AU)} & \colhead{period(days)} & \colhead{Eccentricity} & \colhead{source} & \colhead{$N-star$} 
}
\decimalcolnumbers
\startdata
81 &  GJ 676 e &  1010 &  247.8 &  4 &  0.036 &  0.187 &  35.37 &  0.24 &  2 &  2 \\   
82 &  HD 142022 b &  1030 &  210.6 &  1 &  5.1 &  3.03 &  1928 &  0.53 &  2 &  2 \\ 
83 &  55 Cnc b &  1065 &  350.5 &  5 &  0.8 &  0.1134 &  14.651 &  0.016 &  2 &  2 \\  
84 &  55 Cnc c &  1065 &  350.5 &  5 &  0.169 &  0.2403 &  44.3446 &  0.053 &  2 &  2 \\   
85 &  55 Cnc d &  1065 &  350.5 &  5 &  3.835 &  5.76 &  5218 &  0.025 &  2 &  2 \\   
86 &  55 Cnc f &  1065 &  350.5 &  5 &  0.144 &  0.781 &  260.7 &  0 &  2 &  2 \\ 87 &  16 Cyg b &  1080 &  189.7 &  1 &  1.68 &  1.68 &  799.5 &  0.689 &  2 &  3 \\   
88 &  HD 11964 b &  1272.6 &  258.8 &  2 &  0.622 &  3.16 &  1945 &  0.041 &  2 &  2 \\   
89 &  HD 11964 c &  1272.6 &  258.8 &  2 &  0.079 &  0.229 &  37.91 &  0.3 &  2 &  2 \\   
90 &  HD 101930 b &  2772 &  507.2 &  1 &  0.3 &  0.302 &  70.46 &  0.11 &  2 &  2 \\   
91 &  91 Aqr b &  2835 &  554.8 &  1 &  3.2 &  0.7 &  181.4 &  0.03 &  2 &  3 \\  
92 &  HD 89744 b &  3100 &  856.3 &  2 &  7.2 &  0.88 &  256 &  0.7 &  2 &  2 \\   
93 &  HD 89744 c &  3100 &  856.3 &  2 &  3.2 &  0.44 &  85.2 &  0.29 &  2 &  2 \\  
94 &  HD 190360 b  &  3585.96 &  905.3 &  2 &  1.502 &  3.92 &  2891 &  0.36 &  2 &  2 \\   
95 &  HD 190360 c  &  3585.96 &  905.3 &  2 &  0.057 &  0.128 &  17.1 &  0.01 &  2 &  2 \\   
96 &  HD 213240 b &  4915.378 &  1301.1 &  1 &  4.5 &  2.03 &  951 &  0.45 &  2 &  3 \\   
97 &  XO-2S b &  5796 &  1027 &  2 &  0.259 &  0.1344 &  18.157 &  0.18 &  2 &  2 \\   
98 &  XO-2S c &  5796 &  1027 &  2 &  1.37 &  0.4756 &  120.8 &  0.153 &  2 &  2 \\  
99 &  HD 222582 b &  5990 &  1416.5 &  1 &  7.75 &  1.35 &  572.38 &  0.725 &  2 &  2 \\   
100 &  HD 125612 b &  5990 &  2453.9 &  3 &  3 &  1.37 &  502 &  0.46 &  2 &  2 \\  
101 &  HD 125612 c &  5990 &  2453.9 &  3 &  0.058 &  0.05 &  4.1547 &  0.27 &  2 &  2 \\   
102 &  HD 125612 d &  5990 &  2453.9 &  3 &  7.2 &  4.2 &  3008 &  0.28 &  2 &  2 \\ 103 &  HD 147513 b &  6760 &  1326.9 &  1 &  1.21 &  1.32 &  528.4 &  0.26 &  2 &  2 \\   
104 &  HD 40979 b &  8070.4 &  1764.5 &  1 &  3.28 &  0.83 &  263.1 &  0.25 &  2 &  3 \\   
105 &  Prox Cen b &  10962 &  2041.7 &  2 &  0.003 &  0.0485 &  11.186 &  0 &  2 &  3 \\   
106 &  Prox Cen c &  10962 &  2041.7 &  2 &  0.0182 &  1.48 &  1894 &  0 &  2 &  3 \\   
107 &  HIP 70849 b  &  11340 &  3081.9 &  1 &  9 &  10 &  10000 &  0.6 &  2 &  2 \\  
108 &  HD 20782 b &  11550 &  2386.2 &  1 &  1.9 &  1.381 &  591.9 &  0.97 &  2 &  2 \\   
109 &  HD 38529 c &  15184.962 &  3805.3 &  2 &  17.7 &  3.695 &  2134.76 &  0.36 &  2 &  2 \\   
110 &  HD 38529 b &  15184.962 &  3805.3 &  2 &  0.87 &  0.131 &  14.31 &  0.25 &  2 &  2 \\ 
\enddata
\tablecomments{{\Su$a_B$: the semimajor axis of binary orbit, calculated via the empirical relationship \citep{1992ApJ...396..178F} $a_B$/r = 1.261 (r is projected binary separation) if there is no reported $a_B$ in the source catalog.  $a_c$:the critical distance to the primary star to maintain long-term orbital stability \citep{1999AJ....117..621H}. Source: 1, the catalog of exoplanet-hosting binaries with separations up to 500 AU\footnote{\url{http://exoplanet.eu/planets_binary/}}; 2,  the catalogue of exoplanets in binary star systems \citep{2016MNRAS.460.3598S}\footnote{\url{https://www.univie.ac.at/adg/schwarz/multiple.html}}. N-star:2, binary star system; 3 or 4, triple system or quadruple system.}
\label{tab:tab1}}
\end{deluxetable*}

\end{document}